\documentclass[conference]{ieeeconf}
\IEEEoverridecommandlockouts

\usepackage{amsmath,amssymb,amsfonts,amsthm,bbm}
\usepackage{graphicx}
\usepackage{comment}
\usepackage{textcomp}
\usepackage[makeroom]{cancel}
\usepackage{indentfirst}
\usepackage{ulem}
\usepackage{tikz}
\usepackage{epstopdf} 
\usepackage{epsfig}
\usepackage[dvipsnames]{xcolor}
\usepackage{algorithm,algpseudocode}
\usepackage[linesnumbered,ruled,vlined,algo2e]{algorithm2e}
\usepackage{hyperref}
\usepackage{etoolbox}

\usepackage{optidef}

\newbool{short}

\boolfalse{short} 

\DeclareMathOperator*{\argmin}{arg\,min}

\usepackage{mathtools}
\mathtoolsset{showonlyrefs}

\usepackage[utf8]{inputenc}
\usepackage[english]{babel}
\usepackage[noadjust]{cite}

\renewcommand{\emph}{\textit}

\theoremstyle{definition}
\newtheorem{theorem}{Theorem}

\newtheorem{assumption}{Assumption}
\newtheorem{lemma}{Lemma}
\newtheorem{definition}{Definition}
\newtheorem{problem}{Problem}
\newtheorem{proposition}{Proposition}
\newtheorem{remark}{Remark}

\usepackage{csquotes}
\usepackage{float}
\usepackage{accents}

\def\BibTeX{{\rm B\kern-.05em{\sc i\kern-.025em b}\kern-.08em
    T\kern-.1667em\lower.7ex\hbox{E}\kern-.125emX}}

\newcommand{\R}{\mathbb{R}}
\newcommand{\N}{\mathbb{N}}
\newcommand{\ste}{x}
\newcommand{\Stes}{\zeta}
\newcommand{\inpt}{u}
\newcommand{\outpt}{y}
\newcommand{\soutpt}[1]{y_{s#1}}
\newcommand{\interm}{z}
\newcommand{\atimer}[1]{\tau_{c#1}}
\newcommand{\ctimer}[1]{\tau_{g#1}}

\newcommand{\dist}{d}
\newcommand{\HFO}{\mathcal{H}_{FO}}
\newcommand{\intv}{\tau}
\newcommand{\objpu}[1]{\nabla_{#1}\Phi}

\newcommand{\Htime}[2]{(t_{#1},j_{#2})}
\newcommand{\Hatime}[2]{(t_{#1},j{#2})}
\newcommand{\Hztime}[2]{(t_{#1},{#2})}
\newcommand{\Hotime}[2]{({#1},{#2})}
\newcommand{\alphatime}[1]{\textcolor{black}{\bar{\alpha}\left(#1\right)}}

\newcommand{\HFOR}{\mathcal{H}_{FO}^{\iota}}
\newcommand{\n}[1]{\textnormal{#1}}

\newcommand{\twonorm}[1]{\left\|#1\right\|}

\newcommand{\paren}[1]{\left(#1\right)}

\newcommand{\blue}[1]{\textcolor{black}{#1}} 
\newcommand{\green}[1]{\textcolor{black}{#1}} 
\newcommand{\red}[1]{\textcolor{black}{#1}} 

\usepackage{subfiles} 

\begin{document}
\title{\LARGE \bf A Hybrid Systems Model of Feedback Optimization for \\ Linear Systems: Convergence and Robustness}
\author{
Oscar Jed R. Chuy$^{1}$, Matthew T. Hale$^1$, and Ricardo G. Sanfelice$^2$
    \thanks{
    $^{1}$School of Electrical and Computer Engineering, Georgia Institute of Technology, Atlanta, GA USA. 
    Emails: \texttt{\{ochuy3,matthale\}@gatech.edu}.
    }
    \thanks{
    $^{2}$School of Electrical and Computer Engineering,  University of California, Santa Cruz, CA USA.
    Email: \texttt{ricardo@ucsc.edu}.
    }
    \thanks{
        All authors were supported by AFOSR under grant FA9550-19-1-0169.
        Chuy and Hale were supported by ONR under grants N00014-21-1-2495, N00014-22-1-2435, and N00014-26-1-2068,
        and AFRL under grants FA8651-22-F-1052 and FA8651-23-F-A006. 
        Sanfelice was supported by NSF Grants no. CNS-2039054 and CNS-2111688, by AFOSR Grants nos. FA9550-23-1-0145, FA9550-23-1-0313, and FA9550-23-1-0678, by AFRL Grant nos. FA8651-22-1-0017 and FA8651-23-1-0004, by ARO Grant no. W911NF-20-1-0253, and by DoD Grant no. W911NF-23-1-0158. 
    }
}
\maketitle

\begin{abstract}
Feedback optimization algorithms compute inputs to a system using real-time
output measurements, which helps mitigate the effects of disturbances. 
However, existing work often models both system dynamics and computations in either discrete or continuous time, which may not accurately model some applications.
In this work, we model linear system dynamics in continuous time and the computation of inputs in discrete time, 
\green{leading to} a \green{new} hybrid system model of feedback optimization. 
\green{First, we} establish well-posedness of this hybrid model and establish completeness of solutions while ruling out Zeno behavior.
Then, we show \green{that} the state of the system converges exponentially fast to a ball of known radius about a desired 
goal state. 
Next, we analytically show that this system is robust to perturbations 
\green{over bounded (hybrid) time horizons} in (i) the values of measured outputs, (ii) the 
matrices that model 
the linear time-invariant system, and (iii) the times at which inputs are applied to the system. 
Simulation results confirm that this approach successfully mitigates the effects of disturbances.
\end{abstract}

\section{Introduction} \label{sec:intro}
Many automation tasks require optimizing the behavior of a dynamical system, which often involves solving a planning problem offline. 
With accurate system models, an optimization problem may be solved in a feedforward configuration to generate a reference that is used to drive the system in question~\cite{hauswirth2024optimization,batchRLHanchen2020, processcntrl2010, OPSFfotz2000, KKTcontrollerAndrej, BIGNUCOLORadMV, ORTMANN2020106782}. 
However, errors in a system model 
can lead to sub-optimal solutions~\cite{hauswirth2024optimization}
because the inputs applied to a system may not actually
produce the intended outputs. 

If models are inaccurate, one alternative approach 
called ``feedback optimization''
instead \emph{measures} system outputs~\cite{hauswirth2018stab,hauswirth2024optimization,ORTMANN2020106782,HausT-Sinautoopt,processcntrl2010}
and then uses those measurements to optimize inputs with an in-the-loop optimization algorithm. 
Feedback optimization has been shown to have several benefits in certain settings: it is robust to inaccurate system models and time-varying parameters, achieves constraint satisfaction with minimal model dependence, and eliminates the need for pre-computed set points or reference signals~\cite{hauswirth2024optimization, processcntrl2010}. 
This approach has been used, for example, in 
decentralized settings~\cite{wang2023decentralized, behrendt24}, 
gradient-based feedback control~\cite{HausT-Sinautoopt}, 
zeroth-order optimization~\cite{hauswirth2024optimization},
chemical processes \cite{processcntrl2010}, 
and network congestion control \cite{netflowcontrol1999}. 

In existing work, feedback optimization has usually been applied to systems with 
either (a) continuous-time dynamics and a continuous-time optimization algorithm in the loop or 
(b) discrete-time dynamics and a discrete-time optimization algorithm in the loop~\cite{wang2024online, hauswirth2024optimization, ELLIS20141156}.
However, physical systems are often modeled in continuous time and digital computers are naturally modeled in discrete time, which means that  practical implementations can produce dynamics that are not captured by~(a) or~(b). 

We seek to show that feedback optimization retains its robustness
guarantees with continuous-time dynamics and discrete-time optimization,
and we therefore develop a hybrid systems model of it. 


The contributions of this paper are the following:
\begin{itemize}
    \item We model feedback optimization as a hybrid system and show that (i) it is free from Zeno behavior and (ii) all maximal solutions are complete (Proposition~\ref{prop:maxcomsol}).
    \item We bound the distance between the state of the hybrid feedback optimization model and a desired goal state (Theorem~\ref{th:completeHO} and Theorem~\ref{thm:GCHC}).
    \item We prove robustness of the hybrid model to perturbations by showing that \green{over bounded (hybrid) time horizons} there is bounded difference between the solutions to a perturbed system and the solutions to a nominal system (Theorem~\ref{th:robust}).
    \item We \green{validate with} simulations that hybrid feedback optimization successfully rejects disturbances (Section~\ref{sec:simulations}).
\end{itemize}

Prior work in~\cite{giuseppeOFES, cothren2023perceptionbasedsampleddataoptimizationdynamical,chen25}
combines continuous-time dynamics and 
discrete-time optimization in a sampled-data 
feedback optimization setting,
\green{though we develop new analytical robustness guarantees by developing a hybrid model in the framework of \cite{Hybridbook}.}
The main results in~\cite{giuseppeOFES,cothren2023perceptionbasedsampleddataoptimizationdynamical}
show that the closed-loop system is
stable for large enough sample times and that
it is practically stable under time-varying disturbances
with bounded variation. 
Results in~\cite{chen25}
show global exponential stability of the closed-loop system
when inputs to the system change at a fixed rate \blue{(with periodic sampling times)}.

\green{Our results} 
\green{consider a hybrid system model in the framework}
of~\cite{Hybridbook}, 
\blue{ which allows us to characterize system behavior at all times (permits aperiodic sampling), and unlocks robustness analyses that we apply in this work}. 
We use this model to show that hybrid feedback optimization is simultaneously robust 
to several types of disturbances, 
including disturbances in the values of measured outputs,
errors in the times at which new inputs are applied to the system, and errors in 
all of the matrices that model the linear time-invariant
system that is being controlled.
To the best of our knowledge, this paper is the first to analytically prove that 
a hybrid/sampled-data feedback optimization model
is simultaneously robust to all of these disturbances.

The rest of the paper is organized as follows. 
Section~\ref{sec:problem} provides background. 
Section~\ref{sec:hybridmodelone} gives problem statements and 
the hybrid feedback optimization model. 
Section~\ref{sec:properties} derives properties of its solutions.
Then, 
Section~\ref{sec:convergence} establishes 
the solutions' convergence and the system's robustness. 
Section~\ref{sec:simulations} presents simulations. 
\red{Due to space constraints, proofs will be published elsewhere.}

\section{Preliminaries} \label{sec:problem}
This section gives background on feedback optimization. 
\subsection{Notation}
Let $\R$ denote the reals and let~$\N$ 
denote the non-negative integers. 
For a differentiable function $\Phi\!:\!\R^m\!\times\!\R^p\!\rightarrow\!\R$, let $\objpu{\inpt{}}$ denote the partial derivative with respect to its first argument. 
The symbol $I_b$ denotes the identity matrix of dimension $b$. 
The~$2$-norm of a vector~$x$ is denoted~$\twonorm{\ste}$.
For a non-empty, compact, convex set~$\mathcal{Z}$,
the symbol~$\Pi_{\mathcal{Z}}[v]$ denotes the
Euclidean projection of a point~$v$ onto~$\mathcal{Z}$,
i.e., $\Pi_{\mathcal{Z}}[v] = \argmin_{z \in \mathcal{Z}} \|v - z\|$.
We denote the diameter of a non-empty, compact, convex set
$\mathcal{Z}$ by~$d_{\mathcal{Z}}$. 
We use~$\lambda_i(N)$ to denote the~$i^{th}$ 
eigenvalue of a matrix~$N$,
and we use~$\mathfrak{Re}\{a\}$ to denote the real part of a complex number~$a$. 
We also write~$\lambda_{min}(N)$ 
and~$\lambda_{max}(N)$ for the smallest 
and largest (real) eigenvalues of a symmetric matrix~$N$,
respectively. 
Given~$r\!\geq\!0$, we use~$B_r(\tilde{x})\!:=\!\{x\!\in\!\R^n\!:\!\|x - \tilde{x}\|\!\leq\!r\}$ to denote the closed Euclidean
ball of radius~$r$ about the point $\tilde{x} \in \R^n$. 
For a set~$S$ and a point~$x$, we use~$\|x\|_{S}\!:=\!\inf_{s \in S}\!\|x - s\|$. 
\ifbool{short}{\green{For vector elements~$x$ and~$y$ we denote~$(x^\top \; y^\top)^\top = (x,y)$.}}{}

\subsection{Feedback Optimization Background and Setup} \label{ss:fosetup}
Now we review ``feedback optimization'' as defined in the literature, e.g.,~\cite{hauswirth2024optimization, KRISHNAMOORTHY2022107723}. 
At a high level, this class of problems uses real-time measurements from a dynamical system that are fed into an optimization algorithm in a closed-loop structure. 
The goal in doing so is to 
compute inputs that 
optimize the steady-state behavior of the dynamical system. 

To the best of our knowledge there has not been a systematic investigation of the robustness of hybrid feedback optimization for a continuous-time system driven by discrete-time computations.
Other existing work has developed hybrid models of optimization algorithms~\cite{hendrickson21,hustig23,hendrickson25}, though
we differ by developing a hybrid model of 
feedback optimization.
There has also been work on hybrid model predictive control 
\cite{ALTIN2018128}, but our aim is different because we implement a hybrid framework for feedback optimization.
Work in \cite{giuseppeOFES,cothren2023perceptionbasedsampleddataoptimizationdynamical,chen25} studies feedback optimization in a sampled-data setting, but our use of a hybrid model lets us derive new analytical robustness guarantees that include robustness to errors in the times at which new inputs are applied, perturbations to the values of sampled outputs,
and errors in the plant model.


Suppose we have the linear time-invariant (LTI) system 
\ifbool{short}{
\begin{equation} \label{eq:ltisystem}
    \blue{\dot{\ste} = A\ste + B\inpt \quad\quad
    \outpt = \Psi\ste + \dist,}
\end{equation}
}{
\begin{align} \label{eq:ltisystem}
    \begin{split}
        \dot{\ste} &= A\ste + B\inpt \\
        \outpt &= \Psi\ste + \dist,
    \end{split}
\end{align}
}
where $\ste\in\R^n$ is the system's state, $\inpt \in \mathcal{U} \subset \R^m$ is its input,
$\mathcal{U}$ is a non-empty, compact, convex set, and~$\outpt \in \R^p$ 
is its output. 
The vector $d\in\R^p$ is a constant, unknown disturbance (e.g., bias), 
which is a typical component of feedback optimization problem formulations~\cite{hauswirth2024optimization, wang2023decentralized,Colombino2020feedback ,HausT-Sinautoopt,wang2024online}.
Such disturbances arise for example, in various power systems applications~\cite{hauswirth2024optimization, Colombino2020feedback, HausT-Sinautoopt,wang2023decentralized, wang2024online}. 

\begin{assumption}
    \label{LTI-AS}
    The matrix~$A$ is Hurwitz. 
\end{assumption}
\blue{Without loss of generality, the eigenvalues of~$A$ are ordered such that $\lambda_{n} \leq \lambda_{n-1} \leq \cdots \leq \lambda_1 < 0$.}

\begin{remark}
We adopt the ``stabilize then optimize'' approach~\cite{giuseppeOFES, wang2023decentralized} in which we suppose that a stabilizing
controller has already been applied to the system. 
If Assumption~\ref{LTI-AS} is not satisfied, then it can be enforced for any stabilizable system by 
doing pre-feedback with a stabilizing controller.  
\end{remark}

Assumption~\ref{LTI-AS} 
ensures that~\eqref{eq:ltisystem} will eventually reach steady state \green{when its input is constant},
which allows its steady-state behavior to be optimized. 
From \eqref{eq:ltisystem}, the steady-state input-to-output map is
    $u \mapsto H\inpt + \dist$, where $H:= -\Psi A^{-1}B$. 

\begin{remark} 
    While one could envision using an estimator to determine~$d$, we are interested in cases in which the matrices~$A$,~$B$, and~$\Psi$ are not known exactly, and an observer may have poor accuracy under these conditions. 
    \blue{
    For implementation, exact knowledge of~$H$ is unnecessary as errors are allowed as covered in Section~\ref{subsec:GConvandRobust}.
    }
\end{remark}

To optimize the system's steady-state behavior, one can drive its
input and output to a solution of 
\begin{mini!}|l|
    {\inpt, \outpt{}}
    {\Phi(\inpt,\outpt{})}
    {\label{eq:fo_genform}}
    {\label{eq:phidef}}
    \addConstraint{\outpt = H\inpt + \dist, \,\, \inpt \in \mathcal{U}, \,\, \outpt{} \in \R^p}{\label{eq:ddef} }
\end{mini!}
where 
~$\Phi : \R^m \times \R^p \to \R$ is 
strongly convex in~$(u, y)$. 
These properties ensure~\eqref{eq:phidef}-\eqref{eq:ddef} has a unique solution. 

By incorporating the constraint~$y = Hu + d$ into \eqref{eq:phidef}, one could in principle reduce this problem to
\begin{align}\label{eq:phitilde}
    \underset{u}{\operatorname{minimize}} \quad \tilde{\Phi}(\inpt,  H\inpt + \dist) \quad \operatorname{subject~to} \quad \inpt \in \mathcal{U}
\end{align}
However, the problem  in \eqref{eq:phitilde} cannot be solved in practice 
because the substitution~$\outpt = H\inpt + \dist$ would
require exact knowledge of the disturbance~$\dist$, which may not be available. 
Instead, feedback optimization is used to repeatedly 
measure~$y$ and then optimize over~$u$. 
In this work,~$\outpt$
is sampled at discrete instants of time, and we use~$y_s$ to denote its sampled value. 

\begin{remark} \label{rem:steadystate}
The underlying LTI system need not always
be at steady state, but we will 
use a standard technique in the feedback optimization
literature to approximate its outputs as coming from a system at steady state~\cite{hauswirth2018stab, HausT-Sinautoopt,wang2023decentralized}. 
Mathematically, we 
approximate a sampled output~$\soutpt{}$ as coming
from the steady-state map~$\soutpt{} = H\inpt + \dist$
when the input to the system is~$\inpt$. 
This approximation is 
justified, for example, when the dynamics of the system
converge sufficiently quickly. 
Our results in Sections \ref{sec:properties} and \ref{sec:convergence} still analyze the dynamics for the system, and they use this approximation 
only to relate outputs to inputs. 
\green{Since we optimize over~$u$, we must account for~$u$ when deriving the gradient; however, in its computation we use the sampled value \red{instead}.}
\end{remark}

We study a closed-loop system that connects the LTI
system with a gradient descent algorithm that computes the input~$u$. 
With a sampled output~$y_s$, the optimization update law
is 
    $\inpt_{k+1} = \Pi_{\mathcal{U}}\big[\inpt_{k} -\gamma \nabla_u \Phi(\inpt_k, \soutpt{})\big]$,
where~$u_k$ is the~$k^{th}$ iterate of the gradient descent algorithm. 
Then,
the closed-loop interconnected LTI system is 
\ifbool{short}{
\blue{\begin{align} \label{eq:cdltisystem}
    \text{Plant: }&
    \begin{cases}
        \dot{\ste} = A\ste + B\inpt \quad\quad
        \outpt = \Psi\ste + \dist,
    \end{cases}\\
    \text{Controller:}&
    \begin{cases} 
        \inpt_{k+1} = \Pi_{\mathcal{U}}\big[\inpt_{k} -\gamma\nabla_u\Phi(\inpt_k,\soutpt{})\big].
    \end{cases}
\end{align}}
}{
\begin{align} \label{eq:cdltisystem}
    \text{Plant: }&
    \begin{cases}
        \dot{\ste} &= A\ste + B\inpt \\
        \outpt &= \Psi\ste + \dist,
    \end{cases}\\
    \text{Controller:}&
    \begin{cases} 
        \inpt_{k+1} = \Pi_{\mathcal{U}}\big[\inpt_{k} -\gamma\nabla_u\Phi(\inpt_k,\soutpt{})\big].
    \end{cases}
\end{align}
}


We will formulate and analyze a hybrid model for this interconnection. Neither
the changes in the input nor the measurements of the output will be assumed to 
occur periodically, but instead both can occur sporadically.

\subsection{Background on Hybrid Systems}
In this paper, a hybrid system $\mathcal{H}$ takes the form 
\begin{equation}\label{hybridmodel}
    \mathcal{H} = 
    \begin{cases}
        \begin{aligned}
            \dot{\Stes} &\in F(\Stes) \quad &\Stes \in C \\
            \Stes^+ &\in G(\Stes) \quad &\Stes \in D
        \end{aligned}
    \end{cases},
\end{equation}
where $\Stes \in \R^{n}$ is the system's state vector and the maps~$F$ and~$G$ are set valued in general. 
The \green{set-valued map}~$F$ defines the flow map and governs the continuous dynamics within the flow set $C$, while 
$G$ defines the jump map, which models the system's discrete behavior within the jump set~$D$.

\begin{definition}[Hybrid Basic Conditions \cite{Hybridbook}]
\label{def:hybridcond}
     A hybrid system $\mathcal{H}$ 
    with data~$(C, F, D, G)$ satisfies the hybrid basic conditions if 
    \begin{enumerate}
        \item $C$ and $D$ are closed subsets of $\R^n$; \label{hybridcond_one}
        \item $F: \R^n\rightrightarrows\R^n$ is outer semicontinuous\footnote{A set-valued mapping $M:\R^m\rightrightarrows \R^n$ is outer semicontinuous (osc) at $x \in \R^m$ if  for every sequence of points $\{x_i\}_{i \in \N}$ convergent to $x$ and any convergent sequence of points~$\{y_i\}_{i \in \N}$ with $y_i \in M(x_i)$, one has $y\in  M(x)$, where $\lim_{i\rightarrow\infty}y_i  =  y$ \cite{Hybridbook}.}, and locally bounded\footnote{A set-valued mapping $M:\R^m\rightrightarrows \R^n$ is locally bounded at $x \in \R^m$ if there is a neighborhood $U_x$  of $x$ such that $M(U_x)\subset \R^n$ is bounded \cite{Hybridbook}.} relative to $C$, $C\subset \text{dom}~F,$ and $F(\Stes)$ is convex for every $\Stes \in C$ \label{hybridcond_two}
        \item $G: \R^n\rightrightarrows\R^n$ is outer semicontinuous and locally bounded relative to $D$, and $D\subset \text{dom}~G$. \label{hybridcond_three}
    \end{enumerate}
\end{definition}

If a hybrid system satisfies the hybrid basic conditions, then it is well-posed by \cite[Theorem 6.30]{Hybridbook}. 
We use this property in Section \ref{sec:convergence} to show
that errors in the models of~$F$ and~$G$ up to a certain threshold produce bounded changes in the resulting closed-loop 
system trajectories over compact time horizons. 
It is not automatic to formulate a well-posed hybrid system, and this paper will do so for a hybrid model of feedback optimization. 

For a hybrid system $\mathcal{H}$, its solutions, denoted by~$\phi$, are hybrid arcs that can in general be  maximal\footnote{A solution $\phi$ to $\mathcal{H}$ is maximal if there does not exist another solution $\psi$ to $\mathcal{H}$ such that dom~$\phi$ is a proper subset of dom~$\psi$ and $\phi(t,j)=\psi(t,j)$ for all $(t,j)\in \textnormal{dom } \phi$~\cite{Hybridbook}.}, complete\footnote{The solution $\phi$ is complete if~$ \textnormal{dom }\phi$ is unbounded, i.e., if $\textnormal{length}(\textnormal{dom }\phi) = \sup_t \textnormal{dom }\phi + \sup_j \textnormal{dom }\phi = \infty$~\cite{Hybridbook}.}, and Zeno\footnote{The solution $\phi$ is Zeno if it is complete and $\sup_t \textnormal{dom } \phi<\infty$~\cite{Hybridbook}.}. 
Complete solutions are defined over arbitrarily long time horizons, and
Zeno behavior implies that solutions undergo an infinite number of jumps in finite time. 
Zeno behavior implies that a system's states stop flowing in finite time, which we will rule out
for feedback optimization. 

\section{A Hybrid Model for Feedback Optimization} 
\label{sec:hybridmodelone}

The problems we solve in this paper are as follows. 

\begin{problem} \label{prob:hybrid}
Formulate a well-posed hybrid feedback optimization model of the 
plant and controller in~\eqref{eq:cdltisystem} that captures
(i) how outputs are intermittently measured and used in the optimization-based controller,  and
(ii) how inputs are computed and applied to the system. 
\end{problem}

\begin{problem} \label{prob:convergence}
    \red{Bound the steady-state error of the system relative to a desired goal state in terms of system parameters.}
    
\end{problem}

\begin{problem} \label{prob:robustness}
    Show that the hybrid feedback optimization model is robust to perturbations in the LTI system model, 
    measurements of outputs, and the times at which inputs are applied to the LTI system,
    in the sense that these perturbations induce bounded changes
    in solutions. 
\end{problem}

\subsection{Overview of Hybrid Feedback Optimization}
The continuous-time system in~\eqref{eq:cdltisystem} receives inputs
from the discrete-time optimization algorithm in~\eqref{eq:cdltisystem},
and those inputs only change at certain instants of time. 
Between these changes, 
inputs applied to the system are held constant.  
Similarly, the optimization algorithm measures an output of the system
and uses it to perform some number of computations to optimize inputs. 
This sampled value of the output is held constant while optimizing
an input, and it does not change until a new output is sampled.  

The flow map~$F$ from Definition~\ref{def:hybridcond} 
will model the LTI system dynamics in~\eqref{eq:ltisystem}
with piecewise constant inputs. The jump 
map~$G$ from Definition~\ref{def:hybridcond} will model both
the 
sampling of outputs and the application of new inputs to the system.
It will also model computations done 
by the optimization algorithm, which are generated 
at discrete points in time
between the times at which the output value is sampled. 

\subsection{Optimization Problem Setup}
We consider objectives~$\Phi$ of the form
    \begin{equation}\label{eq:quadOBJ-AS} 
            \green{\Phi(\inpt,\soutpt{})} = \frac{1}{2}u^\top Q_{\inpt{}}u + \frac{1}{2}(\soutpt{} - \hat{y})^\top Q_{\outpt{}}(\soutpt{} - \hat{y}), 
    \end{equation}
    where $Q_{\inpt{}} \in \R^{m \times m}$ and $Q_{\outpt{}} \in \R^{p \times p}$ are symmetric and positive definite, 
    and $\hat{y} \in \mathbb{R}^p$ is a constant, desired output value that is user-specified. 

Quadratic objectives have been widely used in the feedback optimization literature \cite{hauswirth2024optimization, Lawrence2021, wang2023decentralized, wang2024online, behrendt24, KRISHNAMOORTHY2022107723},
and we emphasize that our hybrid model is not restricted to
using quadratic objectives. Instead, in this initial work
we focus on quadratic objectives to develop 
a hybrid systems
model for feedback optimization, and we defer
the study of other
objectives (including nonconvex objectives) to future work that will build on the present paper. 



\subsection{Hybrid Modeling and Flow and Jump Sets}

The state of the hybrid system includes~$\ste \in \R^n$ and~$\inpt \in \R^m$ which are, respectively, the state and input of the LTI system in~\eqref{eq:ltisystem}.
It also includes the vector $\soutpt{} \in \R^p$, which is the value of the sampled output of the LTI system that is used in the underlying optimization algorithm, $\interm \in \R^m$, which is the current iterate of that optimization algorithm, $\atimer{} \in \R$, which is a timer that tracks the amount of continuous time left until the input to the system changes, and $\ctimer{} \in \R$, which is a timer that accounts for the amount of continuous time needed to complete an iteration of the optimization algorithm. The full state is
\ifbool{short}{
\red{$\zeta := 
    \left(
    x, 
    \inpt, 
    \soutpt{}, 
    \interm, 
    \atimer{}, 
    \ctimer{} 
    \right)
    \in \mathcal{X} := \R^{n + 2m + p +2}. $}
}{
\begin{equation} \label{eq:states}
    \zeta := 
    \left(\begin{array}{cccccc}\!
    x^{\top} &
    \inpt^{\top} &
    \soutpt{}^{\top} &
    \interm^{\top} &
    \atimer{} &
    \ctimer{} 
    \end{array}\!\right)^{\top}
    \in \mathcal{X} := \R^{n + 2m + p +2}.
\end{equation}
}

The timers~$\atimer{}$ and~$\ctimer{}$ count down from some positive numbers to zero, and jumps occur only when they reach zero. 
The state is allowed to flow while both~$\atimer{} > 0$ and~$\ctimer{} > 0$, and the state stops flowing and 
undergoes a jump when $\atimer{} = 0$ and/or $\ctimer{} = 0$. 
These conditions are captured by the flow set~$C$ and jump set~$D$, defined as 
\begin{align} 
        C &:= \big\{\Stes \in \mathcal{X} \mid \atimer{} \in [0,\atimer{,\max}], \ctimer{} \in [0,\ctimer{,comp}]\big\} \label{eq:Cdef} \\
        D &:= \big\{\Stes \in \mathcal{X} \mid \atimer{} = 0 \textnormal{ or } \ctimer{} = 0
        \big\}, \label{eq:jpset}
\end{align}
where $\atimer{,max} > 0$ is the maximum amount of time between changes in the 
input and $\ctimer{,comp} > 0$ is the amount of time needed to perform a gradient descent iteration.

\subsection{Flow Map Definition}
The flow map is derived from \eqref{eq:ltisystem}, which defines~$\dot{x}$. 
We note that~$\soutpt{}{}$, the sampled output used by the optimization algorithm, does not vary continuously because it is measured at 
certain time instants and is held constant between measurements. 
The timers~$\atimer{}$ and~$\ctimer{}$ count down to zero continuously and with unit rate, while all other states only change during jumps. Therefore, the flow map is 
\ifbool{short}{
\begin{equation} \label{fwmap}
    \red{F(\Stes) 
    :=
    \left(
    A\ste + B\inpt, 
    0, 
    0, 
    0, 
    {-1}, 
    {-1}
    \right)\forall\zeta\in C.}
\end{equation}
}{
    \begin{equation} \label{fwmap}
        \blue{F(\Stes)} 
        :=
        \left(\begin{array}{c}
        A\ste + B\inpt \\
        0 \\
        0 \\
        0 \\
        -\hspace{-2pt}1 \\
        -\hspace{-2pt}1
        \end{array}\right)
        \quad \textnormal{ for all } \zeta \in C.
    \end{equation}
}


\subsection{Jump Map Definition} \label{ss:jump}
The jump map has three cases: (i) $\ctimer{} = 0$ with~$\atimer{} > 0$, 
(ii) $\atimer{} = 0$ with~$\ctimer{} > 0$, and (iii) $\atimer{} = \ctimer{} = 0$.

In Case~(i), a single gradient descent step has been completed, but since~$\atimer{} > 0$, 
the iterate $\interm$ is not applied as the system input. The jump map for this case 
updates the state $\interm$ using a gradient descent step of the form
$\interm^+ = \Pi_{\mathcal{U}}\big[\interm - \gamma\nabla_u\Phi(\interm, \soutpt{})\big]$, 
where~$\gamma > 0$ is a stepsize.
Here $\soutpt{} \in \R^p$ is the most recently sampled value of the system output. 
The jump map resets $\ctimer{}$ to $\ctimer{,comp}$ so that the computation of a new gradient descent iteration can begin. Other states jump to their current values, which leaves them unchanged.
The jump map for this case is 
\begin{equation} \label{eq:g1def}
   G_1(\Stes) =
            \left(\begin{array}{c}
            \ste \\
            \inpt \\
            \soutpt{} \\
            \Pi_{\mathcal{U}}\big[\interm - \gamma\nabla_u\Phi(z,\soutpt{})\big] \\
            \atimer{} \\
            \ctimer{,comp} 
            \end{array} \right)
            \quad \textnormal{ for all } \zeta \in D_1.
\end{equation}
\blue{where~$D_1 = \{\Stes\in \mathcal{X}: \atimer{} > 0,~\ctimer{} = 0\}$}.

In Case (ii), 
a new input is applied to the system, a new output is sampled, and the timer~$\atimer{}$ resets to some point in the interval~$[\atimer{,min},\atimer{,max}]$, where~$0 < \atimer{,min} \leq  \atimer{,max}$.
This range of times represents indeterminacy in the amount
of time that elapses between the application of successive inputs
to the system. 
When the input changes, it is set equal to~$\interm$. 
When~$\atimer{}$ reaches~$0$, the 
LTI system output is sampled and stored in~$\soutpt{}$ \blue{based on Remark~\ref{rem:steadystate}}, which is held constant until the next sample. 
The states $x$ and~$z$ do not change, and
the jump map is
\begin{equation} \label{eq:g2def}
G_2(\Stes) =
    \left(\begin{array}{c}
            \ste \\
            \interm \\
            Hu + d \\
            \interm \\
            {[\atimer{,min}, \atimer{,max}]} \\
            \ctimer{}
    \end{array} \right)
    \quad \textnormal{ for all } \zeta \in D_2,
\end{equation}
where, \blue{$D_2\!=\!\{\Stes\in \mathcal{X}: \atimer{} = 0,~\ctimer{} > 0\}$} 
and as described in Remark~\ref{rem:steadystate}, 
\blue{we approximate the sampled output~$y_s$ \red{as}~$Hu + d$.}

In Case (iii), 
we combine Cases~(i) and~(ii), and 
the system executes~$G_1$ and then~$G_2$ or~$G_2$ and then~$G_1$. 
The full jump map~$G$ is defined as 
\begin{equation} \label{eq:jpmap}
        G(\Stes) := 
        \begin{cases}
            G_1(\Stes)
            \;\;\; \text{if }
            \atimer{} > 0 \textnormal{ and }\! \ctimer{} = 0 \; \textnormal{Case (i)} \\
            G_2(\Stes)
            \;\;\; \text{if }
            \atimer{} = 0 \textnormal{ and }\! \ctimer{} > 0 \; \textnormal{Case (ii)} \\
            G_3(\Stes) 
           \;\;\; \text{if }
            \atimer{} = 0 \textnormal{ and }\! \ctimer{} = 0 \; \textnormal{Case (iii)}, 
        \end{cases}
\end{equation}
where
$G_3(\Stes) = G_1(\Stes) \cup G_2(\Stes)$. 
Then, the full hybrid model of feedback optimization is
\begin{equation}\label{eq:hybridFO}
    \mathcal{H}_{FO}:= (C,F,D,G),
\end{equation}
where $C$ is from \eqref{eq:Cdef}, $F$ is from \eqref{fwmap}, $D$ is from \eqref{eq:jpset}, and $G$ is from \eqref{eq:jpmap}.

\section{Properties of Hybrid Feedback Optimization} \label{sec:properties}
In this section, we show~$\mathcal{H}_{FO}$ satisfies
certain technical conditions that ensure its 
solutions exist for all time, which completes our solution
to Problem~\ref{prob:hybrid}. 

\subsection{Well-Posedness and Existence of Solutions}
Toward establishing that solutions to~$\mathcal{H}_{FO}$ are defined for all time,
we have the following. 

\begin{lemma} \label{lem:gosc}
The hybrid feedback optimization model~$\mathcal{H}_{FO}$ in~\eqref{eq:hybridFO} is well-posed
in the sense that it satisfies Definition~\ref{def:hybridcond}.
\end{lemma}
\ifbool{short}{
\begin{proof} 
    \red{See authors' technical report~\cite{chuy2025hybridsystemsmodelfeedback}}
\end{proof}
}{\begin{proof}
By inspection, the set~$C$ in~\eqref{eq:Cdef} and the set~$D$ in~\eqref{eq:jpset} together
satisfy
Condition~1 in Definition~\ref{def:hybridcond}. 
The map~$F$ in~\eqref{fwmap} is defined everywhere on~$C$ and outputs a singleton that
is a linear function of the state, 
and thus it satisfies Condition~2 in Definition~\ref{def:hybridcond}.

To show that the map~$G$ in~\eqref{eq:jpmap} satisfies
Condition~3, we can use Lemma~\ref{OSLBcond} in Appendix~\ref{app:gosc}. 
The jump map~$G_1$ in~\eqref{eq:g1def} is outer semicontinuous because the projection mapping~$\Pi_{\mathcal{U}}[\cdot]$
is continuous and because~$G_1$ outputs a singleton. 
The jump map~$G_2$ in~\eqref{eq:g2def} is outer semicontinuous because its only set-valued
entry outputs a compact interval and all other entries are singletons. 
Then the feedback optimization 
jump map in~\eqref{eq:jpmap} has the structure of the jump
map in Lemma~\ref{OSLBcond}, 
and the feedback optimization 
jump set in~\eqref{eq:jpset} 
has the same structure as the jump set in Lemma~\ref{OSLBcond}. 
Therefore, using Lemma~\ref{OSLBcond}, we see that
the feedback optimization jump map~$G$ in~\eqref{eq:jpmap}
is both outer semicontinuous and locally bounded relative to the closed set $D$. 
Then Condition~\ref{hybridcond_three} of Definition \ref{def:hybridcond} is satisfied. Therefore, all conditions
of Definition~\ref{def:hybridcond} are satisfied,
and the system~$\mathcal{H}_{FO}$ is well-posed. 
\end{proof}}

The following result shows that all maximal solutions to the system~$\mathcal{H}_{FO}$ are complete and non-Zeno.

\begin{proposition}[Completeness of Maximal Solutions] \label{prop:maxcomsol}
    Consider the hybrid feedback optimization model~$\mathcal{H}_{FO}$ from~\eqref{eq:hybridFO}. 
    From every point in~$C \cup D$ there exists a nontrivial solution. All maximal solutions
    are complete and non-Zeno.     
\end{proposition}

\ifbool{short}{
\begin{proof}
\red{See authors' technical report~\cite{chuy2025hybridsystemsmodelfeedback}}
\end{proof}

}{
\begin{proof}
    \blue{The high-level steps of the proof include: for a defined tangent cone $T_{C}$, applying Lemma~\ref{lem:complete}, then prove that the only condition possible is that the solutions are complete, and concluded by ruling out Zeno behavior.}
    
    Since $\mathcal{H}_{FO}$ satisfies Definition \ref{def:hybridcond}, we can apply Lemma~\ref{lem:complete} 
    in Appendix~\ref{app:comp}
    to establish the claim.   
    First we will show that condition (VC) holds for all~$\nu \in C \backslash D$. 
    Consider an arbitrary $\nu \in C \backslash D$, and
    let~$U$ be a neighborhood of~$\nu$. 
    We wish to show that $F(\zeta)\cap T_C(\zeta) \neq \emptyset$ for $\zeta\in U\cap C$, where $T_C(\zeta)$ is the tangent cone of the set $C$ at the point $\zeta$.
    Using the flow map~$F$ from~\eqref{fwmap}, we see that only~$\dot{\ste}$, $\dot{\atimer{}}$, and~$\dot{\ctimer{}}$ are non-zero during flows. In addition, $C$ does not restrict $\ste$, which implies that~$x$ can flow in any direction at any time and remain
    feasible. Conversely, the timers~$\atimer{}$ and~$\ctimer{}$ take values in compact intervals, which implies
    that some directions are infeasible at some points in time. 
    Therefore, the satisfaction of the condition~$F(\zeta)\cap T_C(\zeta) \neq \emptyset$ is determined by the dynamics of~$\atimer{}$ and~$\ctimer{}$. 
    
    To show that $F(\zeta)\cap T_C(\zeta) \neq \emptyset$, we compute the tangent cone as
    \begin{align}
        T_{C}(&\Stes) :=~ \\
        &\begin{cases}
            \R^{n+2m+p} \times \{-1\} \times \R   &\text{if $\atimer{} = \atimer{,\max}$} \\ 
            &\hspace{0pt}\text{and $\ctimer{} \in (0, \ctimer{,comp})$}\vspace{5pt}\\
            \R^{n+2m+p} \times \{-1\} \times \{1\} &\text{if $\atimer{} = \atimer{,\max}$}\vspace{-3pt}\\
            &\hspace{0pt}\text{and $\ctimer{} = 0$}\vspace{5pt}\\
            \R^{n+2m+p}\times \{1\} \times \{-1\}  &\text{if 
            $\atimer{} = 0$}\vspace{-3pt}\\
            &\hspace{0 pt}\text{and $\ctimer{} = \ctimer{,comp}$}\vspace{5pt}\\
            \R^{n+2m+p} \times \R \times \{-1\}  &\text{if $\atimer{} \in (\atimer{,\min}, \atimer{,\max})$}\\
            &\hspace{0pt}\text{and $\ctimer{} = \ctimer{,comp}$}\vspace{5pt}\\
            \R^{n+2m+p}\times \{-1\} \times \{-1\} &\text{if $\atimer{} = \atimer{,\max}$ }\vspace{-3pt}\\
            &\hspace{0 pt}\text{and $\ctimer{} = \ctimer{,comp}$}\vspace{5pt}\\
            \R^{n+2m+p} \times \{1\} \times \{1\}  &\text{if $\atimer{} = 0$ }\vspace{-3pt}\\
            &\hspace{0 pt}\text{and $\ctimer{} = 0$}\vspace{5 pt}\\
            \R^{n+2m+p} \times \R \times \R &\text{else.}
        \end{cases}
    \end{align}
    By inspection, it holds that $F(\zeta)\in T_C(\zeta)$ for every~$\zeta \in C \backslash D$ and
    condition~(VC) from Lemma~\ref{lem:complete} is satisfied
    for every~$\nu \in C \backslash D$. Then there exists a nontrivial solution $\phi$ to $\mathcal{H}_{FO}$ with $\phi(0,0) = \nu$. 
    Let~$\mathcal{S}_{\mathcal{H}_{FO}}$ denote the set
    of all such solutions. 
    Every solution $\phi \in \mathcal{S}_{\mathcal{H}_{FO}}$ satisfies one of the three conditions of Lemma~\ref{lem:complete}.

    By inspection, we have~$G(D) \subset C \cup D$, 
    which implies that $3)$ in Lemma~\ref{lem:complete} does not occur. 
    Regarding $2)$, during flows, the components of the solution that change are $\ste, \atimer{}$, and~$\ctimer{}$, and thus only their boundedness needs to be verified. 
    Since $\atimer{}$ and $\ctimer{}$ take values in compact sets, they are bounded and cannot blow up to infinity. 
    And since~$u \in \mathcal{U}$, which is a compact set,
    we see that~$\inpt$ is bounded, and there
    exists some finite~$u_{max} \geq 0$
    such that~$\|\inpt\| \leq u_{max}$.
    Since the mapping~$(x, u) \mapsto Ax + Bu$ is Lipschitz, 
    $F$ in \eqref{fwmap}
    is 
    globally Lipschitz. This property and
    the boundedness of inputs together imply that 
    Condition~$2)$ from Lemma~\ref{lem:complete} does not hold. 
    Therefore, Condition~$1)$ from Lemma~\ref{lem:complete} 
    does hold, and all maximal solutions to~$\mathcal{H}_{FO}$
    are complete. 
    
    Resets {in}~$G$ are triggered only when
    either one of the timers has reached zero, and $G$ resets
    any timers that have reached zero to non-zero values.
    Then~$G(D) \cap D = \emptyset$, which
    rules out Zeno behavior 
    by Proposition~2.34 in~\cite{hybridfeedcntrl}. 
    Then all maximal solutions are complete
    and non-Zeno. 
\end{proof}}

\remark{
\blue{
    Even though Case (iii) in~\eqref{eq:jpmap} introduces nondeterminism (whether $G_1$ or $G_2$ is first), this is an intended property that only applies two \green{consecutive} jumps (\green{hence,} Zeno \green{is not possible}). 
}
}


\subsection{Algorithm Framework} \label{ss:alg}
\blue{
We impose the following assumption about the computation of inputs and the value of the inputs applied.
\begin{assumption} \label{as:timescale}
    There exists~$\ell \in \mathbb{N}$ with~$\ell \geq 1$
    such that~$\ell\ctimer{,comp} \leq \atimer{,\min}$. 
\end{assumption}
Assumption~\ref{as:timescale} ensures that, when the system is properly initialized, at least~$\ell$ gradient descent iterations are performed between any consecutive changes in the input. This condition is a mild form of timescale separation and it will be used later in 
Theorem \ref{th:completeHO}. 
}

\blue{
Consider an initial condition~$\phi\Hotime{0}{0} = \nu \in \mathcal{X}$
that satisfies
\begin{multline} \label{eq:initconds}
\atimer{}(0,0) \in [\atimer{,\min},\atimer{,\max}],
\ctimer{}(0, 0) = \tau_{g,comp}, \textnormal{ and } \\
z\Hotime{0}{0} = u\Hotime{0}{0},
\end{multline}
and  consider
a solution~$\phi$ to~$\mathcal{H}_{FO}$ with those initial conditions.}
\red{With regards to the state component~$z$, we add a subscript to help denote how many gradient descent
iterations there have been in computing the next value of the input~$u$.}
\red{More generally, we} use~$\alpha(i)$ to denote the number of \red{iterates} that are performed when computing the~$(i+1)^{th}$ value
of the input~$u$. 
Assumption~\ref{as:timescale} 
implies that~$\alpha(i) \geq \ell \geq 1$ for all~$i \in \mathbb{N}$.

\blue{At the \green{initial} hybrid time $\Hotime{0}{0}$, the}
initial input to the system~$u(0, 0)$ is applied \blue{and held constant.} The system then performs~$\alpha(0)$ gradient descent iterations (which are~$\alpha(0)$ Case~(i) jumps) before the input to the LTI system is changed.
\blue{When a Case~(ii) jump occurs and is the~$(\alpha(0)  + 1)^{th}$ jump, a new input is applied~$u\Hztime{\alpha(0) + 1}{\alpha(0) + 1}$ where~$\alpha(0)$ Case (i) jumps and \green{one} Case (ii) jump have occurred. 
When computing~$u\Hztime{\alpha(0) + 1}{\alpha(0) + 1}$, the~$k^{th}$ such iterate is denoted~$z_k(t_k, k)$ for any $k\in\{0,1,\cdots,\alpha(0)\}$. For computing~$u\Hztime{\alpha(0) + 1}{\alpha(0) + 1}$, we denote the last iterate $\alpha(0)$ by $z_{\alpha(0)}\Hztime{\alpha(0)}{\alpha(0)}$.}


\begin{figure}
    \centering
    \includegraphics[width=0.9\linewidth]{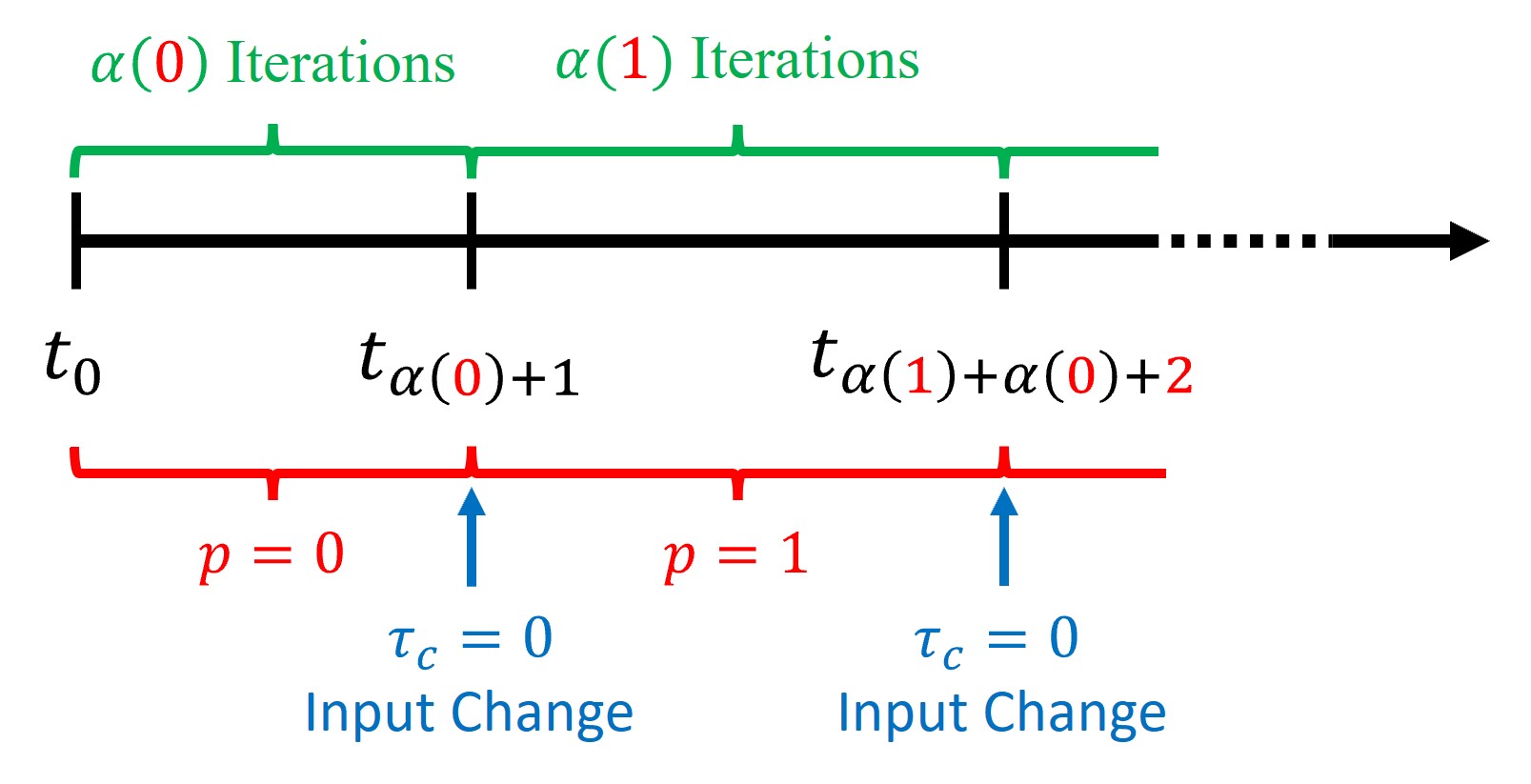}
    \caption{A visual representation of the input evolution of $\mathcal{H}_{FO}$. There are $\alpha(p)$ iterations of gradient descent when computing the $(p+1)^{\textnormal{th}}$ input. The input is changed at hybrid times of the form $(t_{\alphatime{p}+p},\alphatime{p}+p)$.} 
    
    \label{fig:hybridtime}
\end{figure}

\blue{
When $\atimer{}$ reaches zero, a Case (ii) jump occurs and is modeled in~$G_2$ in~\eqref{eq:g2def}. 
The input~$u$ is then set equal to the most recent optimization iterate, i.e.
$u(t_{\alpha(0)+1}, \alpha(0)+1) = z_{\alpha(0)}\big(t_{\alpha(0)}, \alpha(0)\big)$,
and that iterate is used as the next initial iterate for computing the next input so~$\interm_{0}\Hztime{\alpha(0) + 1}{\alpha(0) + 1} = \interm_{\alpha(0)}\Hztime{\alpha(0)}{\alpha(0)}$,
}
i.e., the initial iterate when computing the next input is set
equal to the final iterate that was generated when computing the previous input. 
\blue{For a generalized case with~$p \in \mathbb{N}$, we use~$\alpha(p)$ to denote
the number of gradient descent iterations
that are generated when computing the~$(p+1)^{\n{th}}$ input
to the system. 
We denote \green{by}~$\alphatime{p}$ the total number of gradient
descent iterations that have been computed 
for any input
up until the~$p^{th}$ jump in~$u$. }
That is, $\alphatime{0} = 0$ and
$\alphatime{p} = \sum_{i=0}^{p-1} \alpha(i)$. 
Figure~\ref{fig:hybridtime} illustrates changes in inputs \blue{for the first two jumps.} 

We can identify the general pattern that occurs at an arbitrary Case~(i) jump.
Suppose that~$p$ total Case~(ii) jumps have occurred so far, which means that the input has changed $p$ times. 
Then, when a Case~(i) jump occurs, 
it computes an optimization iterate of the form
\begin{multline} \label{eq:ziterations}
z_{k+1}\Hztime{\alphatime{p} + p + k+1}{\alphatime{p} + p + k+1} = \\
\Pi_{\mathcal{U}}\big[
z_{k}\Hztime{\alphatime{p} + p + k}{\alphatime{p} + p + k} \\
- \gamma\nabla_u\Phi\big(z_{k}\Hztime{\alphatime{p} + p + k}{\alphatime{p} + p + k} \big)\big],
\end{multline}
where for notational simplicity we have used
\begin{multline}
    \nabla_u\Phi\big(z_{k}\Hztime{\alphatime{p} + p + k}{\alphatime{p} + p + k} \big) = \\
    Q_uz_{k}\Hztime{\alphatime{p} + p + k}{\alphatime{p} + p + k} \\
     + H^\top Q_{y}\big(y_s\Hztime{\alphatime{p} + p + k}{\alphatime{p} + p + k} - \hat{y} \big),
\end{multline}
\blue{which by Remark \ref{rem:steadystate} \green{enforces} how, for the optimization problem,
\green{we optimized over~$u$ but sampled~$y_s$ rather than computing it.}}
\blue{The hybrid time $\Hztime{\alphatime{p} + p + k}{\alphatime{p} + p + k}$ accounts for~$\alphatime{p}$ total optimization iterations that were computed up to the~$p^{th}$ }
change in the input, $p$ changes in the input,
and~$k$ optimization iterations that have been computed for the~\blue{$(p+1)^{th}$} input. 

For a Case~(ii) jump,
the~$(p+1)^{th}$ input to the system is 
\begin{multline} \label{eq:c2input}
    \blue{\inpt(t_{\alphatime{p+1}+p+1}, \alphatime{p+1} + p+1)} = \\
    \interm_{\alpha(p)}(t_{\alphatime{p} + \alpha(p) + p},
    \alphatime{p} + \alpha(p) + p),
\end{multline}
which indicates that the new value of the input is equal to the most recently computed optimization iterate. 
The initial iterate for computing the \blue{$(p+2)^{th}$} input is~\blue{$\interm_0(t_{\alphatime{p+1} + p+1}, \alphatime{p+1} + p+1)$},
and it is set equal to the same value as~\blue{$\inpt(t_{\alphatime{p+1}+p+1}, \alphatime{p+1} + p+1)$}.

\section{Convergence Analysis} \label{sec:convergence}
This section solves Problem~\ref{prob:convergence} and
bounds the distance from the
state of the LTI system to its optimal steady-state value. 
The optimization problem in~\eqref{eq:fo_genform}
with an objective of the form of~\eqref{eq:quadOBJ-AS}
has a solution that we denote~$(\tilde{u}, \tilde{y})$, and this solution
depends on the unknown disturbance~$d$
and the constant reference~$\hat{y}$. The optimal
steady-state value of~$x$ is~$\tilde{x} = -A^{-1}B\tilde{u}$ (where~$A^{-1}$
exists under Assumption~\ref{LTI-AS}), and 
thus~$\tilde{x}$ 
incorporates the effects of~$d$ and~$\hat{y}$. 

\ifbool{short}{}{\subsection{States with Piecewise Constant Inputs}
The next result computes the difference between the state~$x$
and its optimal steady-state value~$\tilde{x}$.

\begin{lemma} \label{lem:jpexconvg}
    Consider the hybrid system~$\mathcal{H}_{FO}$
    from~\eqref{eq:hybridFO} and
    suppose that Assumption \ref{LTI-AS} holds. 
    Consider objectives of the form of~\eqref{eq:quadOBJ-AS}. 
    Let~$\phi$ be a maximal solution to~$\mathcal{H}_{FO}$ with initial condition~$\phi(0, 0) = \nu$ that satisfies~\eqref{eq:initconds},
    and consider a hybrid time~$(t, j) \in \textnormal{dom } \phi$.  
    Let~$(\tilde{u}, \tilde{y})$ denote the solution
    to~\eqref{eq:fo_genform} and define~$\tilde{x} = -A^{-1}B\tilde{u}$. 
    Then
    \begin{multline}\label{eq:trackerror_jump}
        \ste(t, j) - \tilde{x}        
         = e^{At}\ste\Hotime{0}{0} \\
        + \sum^{P-1}_{p=0}\int_{t_{\alphatime{p}+p}}^{t_{\alphatime{p+1}+p+1}} e^{A(t_{\alphatime{p+1}+p+1}-\intv)} 
        d\intv \\ \cdot B\inpt\Hztime{\alphatime{p}+p}{\alphatime{p}+p}  \\
        + \int_{t_{\bar{\alpha}(P) + P}}^{t} e^{A(t - \intv)} d\intv \cdot B\inpt\Hztime{\alphatime{P}+P}{\alphatime{P}+P}
        - \tilde{x},  
    \end{multline}
    where we define~$t_0 := 0$, the integer $P = \max\{p \in \mathbb{N} : \bar{\alpha}(p) + p \leq j\}$ is the number of times the input
    to the LTI system has changed before hybrid time~$(t, j)$,
    and~$x$ is \red{the $x$ state} component of the \red{solution}~$\phi$. 
\end{lemma}

\begin{proof}
    The model of~$\mathcal{H}_{FO}$ applies piecewise constant inputs to the underlying LTI system. 
    We will therefore integrate the underlying LTI dynamics over the intervals
    across which the input is constant and 
    compute the difference between~$x$ and~$\tilde{x}$.
    For~$p \in \mathbb{N}$, 
    the input is equal to~$u(t_{\bar{\alpha}(p) + p}, \bar{\alpha}(p) + p)$ 
    over intervals of the form~$[t_{\bar{\alpha}(p) + p}, t_{\bar{\alpha}(p+1) + p+1}]$, 
    \red{over which our result follows.}
\end{proof}}
\ifbool{short}{}{\subsection{Input Convergence}
The following lemma
relates successive iterates that are used to compute the inputs
to the LTI system. In it, we use
\begin{equation} \label{eq:L}
L = \lambda_{max}(Q_u + H^{\top}Q_yH). 
\end{equation}

\begin{lemma}[Input Convergence Rate] \label{lem:inconvrate}
    Consider the hybrid system~$\mathcal{H}_{FO}$
    from~\eqref{eq:hybridFO} and
    consider an objective of the form of~\eqref{eq:quadOBJ-AS}. 
    Suppose that the gradient descent algorithm uses
    a stepsize~$\gamma \in \paren{0,\frac{2}{\lambda_{min}(Q_u) + L}}$. 
    Let~$\phi$ denote a maximal solution to~$\mathcal{H}_{FO}$
    with initial condition~$\phi(0, 0) = \nu$ that satisfies~\eqref{eq:initconds}.
    For any~$(t, j) \in \textnormal{dom } \phi$, 
    set~$P = \max\{p \in \mathbb{N} : \bar{\alpha}(p) + p \leq j\}$. 
    Then, for any integer~$p \in \{0, \ldots, P\}$ 
    the state \red{component}~$z$ of\red{~$\phi$} obeys
    \begin{multline} \label{eq:inputbound}
        \left\|\interm_{\alpha({p})}\Hztime{\alphatime{p} + \alpha({p}) + p}{\alphatime{p} + \alpha({p}) + p}\right.\\
        \left.-\interm^*\Hztime{\alphatime{p} + p}{\alphatime{p} + p}\right\| \\ 
        \leq q^{\frac{\alpha({p})-1}{2}}\left\|\interm_{1}\Hztime{\alphatime{p} + 1 + p}{\alphatime{p} + 1 + p} \right.\\
        \left.- \interm^*\Hztime{\alphatime{p} + p}{\alphatime{p} + p}\right\|,
    \end{multline}
    where 
    \begin{equation}
        z^*\Hztime{\alphatime{p} + p}{\alphatime{p} + p} = \argmin_{u \in \mathcal{U}} \Phi\big(u, \soutpt{}(t_{\alphatime{p}+p}, \alphatime{p}+p)\big)
    \end{equation}
    and $q:= 1-2\gamma\lambda_{min}(Q_u) + \gamma^2L^2\in \paren{0,1},$
    where~$L$ is from~\eqref{eq:L}. 
\end{lemma}

\begin{proof}
    The steps of the proof follow those of a standard proof in the convex optimization literature for the minimization of a strongly convex function, though we present the proof in the hybrid context.
    For the hybrid system $\HFO$, we can quantify convergence of 
    the computation of inputs by examining the distance
    of an intermediate iterate from its optimal value
    after~$k+1$ iterations of gradient descent. 
    That is, we can bound the term
    \begin{multline} 
        \left\|\interm_{k+1}\Hztime{\alphatime{p} + k + 1 + p}{\alphatime{p} + k + 1 + p}\right.\\
        \left.-\interm^*\Hztime{\alphatime{p} + p}{\alphatime{p} + p}\right\|^2,
    \end{multline}
    where~$k+1$ is some number of iterations
    between~$1$ and~$\alpha({p})$. 
    We can then express $\interm_{k+1}$ 
    in terms of~$z_k$ and use the fact that
    \begin{multline}
            \interm^*\Hztime{\alphatime{p} + p}{\alphatime{p} + p} =
            \Pi_{\mathcal{U}}\Big[\interm^*\Hztime{\alphatime{p} + p}{\alphatime{p} + p}
            \\ - \gamma\nabla_u\Phi\big(\interm^*\Hztime{\alphatime{p} + p}{\alphatime{p} + p}, y_s(t_{\alphatime{p} + p}, \alphatime{p} + p)\big)
            \Big],
    \end{multline}
    i.e., $\interm^*\Hztime{\alphatime{p} + p}{\alphatime{p} + p}$ 
    is a fixed point of the projected gradient descent update law. 
    Doing so gives
    \begin{multline} 
        \left\|\interm_{k+1}\Hztime{\alphatime{p} + k + 1 + p}{\alphatime{p} + k + 1 + p}\right.\\
        \left.-\interm^*\Hztime{\alphatime{p} + p}{\alphatime{p} + p}\right\|^2\\ 
        = \left\|\Pi_{\mathcal{U}}\big[\interm_{k}\Hztime{\alphatime{p} + k + p}{\alphatime{p} + k + p}\right.\\
        \left.-\gamma\nabla_{\inpt}\Phi\paren{\interm_{k}\Hztime{\alphatime{p} + k + p}{\alphatime{p} + k + p}}\big]\right.\\
        -\Pi_{\mathcal{U}}\big[\interm^*\Hztime{\alphatime{p} + p}{\alphatime{p} + p} \\
        \left.-\gamma\nabla_{\inpt}\Phi\paren{\interm^*\Hztime{\alphatime{p} + p}{\alphatime{p} + p}}\big]\right\|^2, \label{eq:gdreadytoexpand}
    \end{multline}
    where for ease of notation we have used
    \begin{multline*}
        \nabla_{\inpt}\Phi\paren{\interm_{k}\Hztime{\alphatime{p} + k + p}{\alphatime{p} + k + p}}:= \\\nabla_{\inpt}\Phi(\interm_{k}\Hztime{\alphatime{p} + k + p}{\alphatime{p} + k + p}, \soutpt{}\Hztime{\alphatime{p} + p}{\alphatime{p} + p}),
    \end{multline*}
    and similar for~$\nabla_{\inpt}\Phi\paren{\interm^*\Hztime{\alphatime{p} + p}{\alphatime{p} + p}}$. 
    The non-expansive property of~$\Pi_{\mathcal{U}}$ lets us remove the projections
    and attain an upper bound. Doing this and expanding gives
    \begin{multline}
        \left\|\interm_{k+1}\Hztime{\alphatime{p} + k + 1 + p}{\alphatime{p} + k + 1 + p}\right.\\
        \left.-\interm^*\Hztime{\alphatime{p} + p}{\alphatime{p} + p}\right\|^2\\
        \leq \left\|\interm_{k}\Hztime{\alphatime{p} + k + p}{\alphatime{p} + k + p} \right.\\
        \left.-\interm^*\Hztime{\alphatime{p} + p}{\alphatime{p} + p}\right\|^2\\
        -2\gamma\left(\interm_{k}\Hztime{\alphatime{p} + k + p}{\alphatime{p} + k + p} \right.\\
        \left.-\interm^*\Hztime{\alphatime{p} + p}{\alphatime{p} + p}\right)^\top \cdot \\
        \left(\nabla_{\inpt}\Phi\paren{\interm_{k}\Hztime{\alphatime{p} + k + p}{\alphatime{p} + k + p}} \right.\\
        \left.- \nabla_{\inpt}\Phi\paren{\interm^*\Hztime{\alphatime{p} + p}{\alphatime{p} + p}}\right)\\
        + \gamma^2\left\|\nabla_{\inpt}\Phi\paren{\interm_{k}\Hztime{\alphatime{p} + k + p}{\alphatime{p} + k + p}} \right.\\
        \left.- \nabla_{\inpt}\Phi\paren{\interm^*\Hztime{\alphatime{p} + p}{\alphatime{p} + p}}\right\|^2.
    \end{multline}
    Using the~$L$-Lipschitz property of~$\nabla \Phi_u$
    from~\eqref{eq:L}, we find 
    \begin{multline}
        \left\|\interm_{k+1}\Hztime{\alphatime{p} + k + 1 + p}{\alphatime{p} + k + 1 + p}\right.\\
        \left.-\interm^*\Hztime{\alphatime{p} + p}{\alphatime{p} + p}\right\|^2 \\
        \leq \left\|\interm_{k}\Hztime{\alphatime{p} + k + p}{\alphatime{p} + k + p} \right.\\
        \left.-\interm^*\Hztime{\alphatime{p} + p}{\alphatime{p} + p}\right\|^2\\
        -2\gamma\left(\interm_{k}\Hztime{\alphatime{p} + k + p}{\alphatime{p} + k + p} \right.\\
        \left.-\interm^*\Hztime{\alphatime{p} + p}{\alphatime{p} + p}\right)^\top \cdot \\
        \left(\nabla_{\inpt}\Phi\paren{\interm_{k}\Hztime{\alphatime{p} + k + p}{\alphatime{p} + k + p}} \right.\\
        \left.- \nabla_{\inpt}\Phi\paren{\interm^*\Hztime{\alphatime{p} + p}{\alphatime{p} + p}}\right)\\
        + \gamma^2L^2\left\|\interm_{k}\Hztime{\alphatime{p} + k + p}{\alphatime{p} + k + p} \right.\\
        \left.- \interm^*\Hztime{\alphatime{p} + p}{\alphatime{p} + p}\right\|^2.
    \end{multline}
    Then, since~$Q_u$ is symmetric and positive definite, the function~$\Phi(\cdot, \soutpt{}\Hztime{\alphatime{p} + p}{\alphatime{p} + p})$ is~$\lambda_{min}(Q_u)$-strongly
    convex, and its gradient is~$\lambda_{min}(Q_u)$-strongly monotone. 
    Then we find 
    \begin{multline}
        \left\|\interm_{k+1}\Hztime{\alphatime{p} + k + 1 + p}{\alphatime{p} + k + 1 + p}\right.\\
        \left.-\interm^*\Hztime{\alphatime{p} + p}{\alphatime{p} + p}\right\|^2 \\
        \leq \left\|\interm_{k}\Hztime{\alphatime{p} + k + p}{\alphatime{p} + k + p} \right.\\
        \left.-\interm^*\Hztime{\alphatime{p} + p}{\alphatime{p} + p}\right\|^2\\
        -2\gamma\lambda_{min}(Q_u)\left\|\interm_{k}\Hztime{\alphatime{p} + k + p}{\alphatime{p} + k + p} \right.\\
        \left.- \interm^*\Hztime{\alphatime{p} + p}{\alphatime{p} + p}\right\|^2\\
        + \gamma^2L^2\left\|\interm_{k}\Hztime{\alphatime{p} + k + p}{\alphatime{p} + k + p} \right.\\
        \left.- \interm^*\Hztime{\alphatime{p} + p}{\alphatime{p} + p}\right\|^2,
    \end{multline}
    which simplifies to 
    \begin{multline}
        \left\|\interm_{k+1}\Hztime{\alphatime{p} + k + 1 + p}{\alphatime{p} + k + 1 + p}\right.\\
        \left.-\interm^*\Hztime{\alphatime{p} + p}{\alphatime{p} + p}\right\|^2 \\ 
        \leq \paren{1-2\gamma\lambda_{min}(Q_u) + \gamma^2L^2}\left\|\interm_{k}\Hztime{\alphatime{p} + k + p}{\alphatime{p} + k + p} \right.\\
        \left.- \interm^*\Hztime{\alphatime{p} + p}{\alphatime{p} + p}\right\|^2\\
        = q\left\|\interm_{k}\Hztime{\alphatime{p} + k + p}{\alphatime{p} + k + p} \right.\\
        \left.- \interm^*\Hztime{\alphatime{p} + p}{\alphatime{p} + p}\right\|^2. \label{eq:qbound}
    \end{multline}
    To ensure that $q \in \paren{0,1}$, we use~\cite[Theorem 2.1.15]{convexNetserov}, which shows that~$\gamma \in \paren{0,\frac{2}{\lambda_{min}(Q_u)+L}}$ gives~$q \in (0, 1)$. Then iteratively applying~\eqref{eq:qbound}
    and taking the square root completes the proof. 
\end{proof}

In Lemma \ref{lem:inconvrate} the term $q^{\frac{\alpha({p})-1}{2}}$ shows why the condition $\ell\ctimer{,comp} \leq \atimer{,min}$ is required in Assumption~\ref{as:timescale}. 
If we did not enforce that condition, then the exponent
of~$q$ could be negative, in which case 
the optimization algorithm could diverge. 
}

\subsection{Complete Hybrid Convergence}
The next result is our first main result. It bounds the distance from the 
state of the underlying LTI system, namely~$x$, to its optimal 
steady-state value,~$\tilde{x}$. 
To state this result, we define 
$\mathcal{Y} = \{y \in \mathbb{R}^p : y = Hu + d, u \in \mathcal{U}\}$,
which is compact due to~$\mathcal{U}$ being compact. 
Mathematically, we bound the distance 
between a solution~$\phi$ of~$\mathcal{H}_{FO}$ and the set 
\begin{equation} \label{eq:closedsetA}
\mathcal{A} := {B_r(\tilde{x}}) \times \mathcal{U} \times \mathcal{Y} \times \mathcal{U} \times [0, \atimer{,\max}] \times [0, \ctimer{,comp}],
\end{equation}
where 
$r = M\twonorm{B}d_{\mathcal{U}}\rho^{-1}\big(2 - \exp(-\rho\atimer{,\min})+ q^{\frac{\ell}{2}}\big)$,
$M \geq 1$ is a constant,  
$q:= 1-2\gamma\lambda_{min}(Q_u) + \gamma^2L^2\in \paren{0,1}$,
\begin{equation} \label{eq:rhodef}
\rho = \min_{i \in \{1, \ldots, n\}} |\mathfrak{Re}\{\lambda_i(A)\}|,
\end{equation}
and~$\tilde{x} = -A^{-1}B\tilde{u}$, where~$(\tilde{u}, \tilde{y})$
is the solution to~\eqref{eq:phidef}-\eqref{eq:ddef}.
By definition, we have~$\|\phi(t, j)\|_{\mathcal{A}} = \|x\red{(t, j)}\|_{B_r(\tilde{x})}$. 
Hence, bounding~$(t, j) \mapsto \|\phi(t, j)\|_{\mathcal{A}}$
for each solution to~$\HFO$ allows us to characterize
the error between~$x$ and~$\tilde{x}$
while accounting
for the full dynamics of~$\HFO$. 


\begin{theorem}[Complete Hybrid Convergence] \label{th:completeHO}
    Consider the hybrid system~$\HFO$
    from~\eqref{eq:hybridFO} and suppose that  
    Assumptions~\ref{LTI-AS} and~\ref{as:timescale} hold.
    Consider objectives of the form of~\eqref{eq:quadOBJ-AS}, 
    and suppose that the gradient descent algorithm uses
    a stepsize~$\gamma \in \big(0,\frac{2}{\lambda_{min}(Q_u) + L}\big)$,
    where~$Q_u$ is from~\eqref{eq:quadOBJ-AS} and~$L := \lambda_{max}(Q_u + H^{\top}Q_yH)$. 
    For each maximal solution~$\phi$ to~$\mathcal{H}_{FO}$ 
    with initial condition~$\phi(0, 0) = \nu$ that satisfies~\eqref{eq:initconds}, 
    for each~$(t, j) \in \textnormal{dom } \phi$,      
    \begin{multline}\label{eq:trackerror_complete}
        \twonorm{\phi(t, j)}_{\mathcal{A}} \leq M\exp({-\rho} t) \|\phi(0, 0)\|_{\mathcal{A}} \\
        +\frac{M^2\twonorm{B}d_{\mathcal{U}}}{\rho}\Big(2 - \exp(-\rho\atimer{,\max}) + q^{\frac{\ell}{2}}\Big)\exp({-\rho} t)\\
        - \frac{M\twonorm{B}d_{\mathcal{U}}}{\rho}\Big(1 + q^{\frac{\ell}{2}}\exp({\rho}\atimer{,\min})\Big)\exp({-\rho} t)
   \end{multline}
   where
    $d_{\mathcal{U}} = \max_{u_1, u_2 \in \mathcal{U}} \|u_1 - u_2\|$ is the diameter
    of the set~$\mathcal{U}$, $\rho$ is from~\eqref{eq:rhodef}, 
    $\ell \geq 1$ is from Assumption~\ref{as:timescale},
    $\mathcal{A}$ is from~\eqref{eq:closedsetA}, 
    \green{$q:= 1-2\gamma\lambda_{min}(Q_u) + \gamma^2L^2\in \paren{0,1}$}, 
    and~$M \geq 1$ is a constant.   
    In particular, each such solution satisfies 
        $\lim_{t+j\rightarrow\infty}\twonorm{\phi\Htime{}{}}_{\mathcal{A}} = 0$.
\end{theorem}

\ifbool{short}{
\begin{proof} 
\red{See authors' technical report~\cite{chuy2025hybridsystemsmodelfeedback}}
\end{proof}
}{
\begin{proof}
    \blue{
    In this proof, we note that any empty sums evaluate to zero, and
    we take $\alpha(-1) = 0$ for notational convenience. 
    \red{For each maximal solution~$\phi$ to~$\HFO$,} the high-level steps of the proof are:
    \begin{enumerate}
        \item Bound the distance between the input $u$ and the optimal input $u^*$. \label{step:1}
        \item Express the difference~$x\Hatime{}{} - \tilde{x}$
        in terms of integrals of the dynamics of~$x(t, j)$. \label{step:2}
        \item Evaluate the resulting integral and simplify to bound~$\|x\Hatime{}{} - \tilde{x}\|$. \label{step:3}
    \end{enumerate}
    }
       \blue{We begin with Step~\ref{step:1}.}
    By definition of~$\mathcal{A}$, only the state \red{component}~$x$ in~$\phi$ affects the
    value of~$\|\phi(t, j)\|_{\mathcal{A}}$. 
    That is, we have
    \begin{equation} \label{eq:t1normsequal}
        \|\phi(t, j)\|_{\mathcal{A}} 
        = \|x(t, j)\|_{B_r(\tilde{x})},
    \end{equation}
    and we therefore focus our analysis on~$x$. 
    We define~$P = \max\{p \in \mathbb{N} : \bar{\alpha}(p) + p \leq j\}$.
    Using \eqref{eq:c2input}, we observe that for each~$p \in \{0,\red{1}, \ldots, P\}$ we have 
    \begin{multline}
        \inpt{}\Hztime{\alphatime{p}+p}{\alphatime{p}+p} \\ 
        = z_{\alpha({p-1})}\big(t_{\alphatime{p-1} + \alpha({p-1}) + p-1}, \\ 
            \alphatime{p-1} + \alpha({p-1}) + p-1\big). 
    \end{multline}
    The iterates that are computed to give that input are working towards the optimizer
    \begin{equation}
        \inpt^*{}\Hztime{\alphatime{p}+p}{\alphatime{p}+p} = z^*\Big(t_{\alphatime{p-1} + p-1}, \\ 
            \alphatime{p-1} + p-1\Big),
    \end{equation}
    which is defined as
    \begin{multline}
        \inpt^*{}\Hztime{\alphatime{p}+p}{\alphatime{p}+p} 
        \\= \argmin_{u \in \mathcal{U}} \Phi\Big(u, \soutpt{}\Hztime{\alphatime{p-1} + p-1}{\alphatime{p-1} + p-1}\Big). 
    \end{multline}
    We then find that    
    \begin{multline}
        \twonorm{\inpt{}\Hztime{\alphatime{p}+p}{\alphatime{p}+p} - \inpt{}^*\Hztime{\alphatime{p}+p}{\alphatime{p}+p}}\\
        \leq  
        q^{\frac{\alpha({p-1})-1}{2}}
        \left\|
        z_0\big(t_{\alphatime{p-1} + p-1}, \alphatime{p-1} + p-1\big) \right. \\ 
        -\gamma(Q_u + H^{\top}Q_yH)z_0\big(t_{\alphatime{p-1} + p-1}, \alphatime{p-1} + p-1\big) \\
        \left. - \gamma H^\top Q_y (d - \hat{y}) - \interm^*\big(t_{\alphatime{p-1} + p-1}, \alphatime{p-1} + p-1\big)
        \right.
        \\ 
        \left. + \gamma\big(Q_u + H^TQ_yH\big)\interm^*\big(t_{\alphatime{p-1} + p-1}, \alphatime{p-1} + p-1\big)
        \right.
        \\
        \left.
        + \gamma H^TQ_y(d - \hat{y})
        \right\|, \label{eq:t1reallybigbound}
    \end{multline}
    where we have taken the following steps.
    First, we have applied Lemma~\ref{lem:inconvrate}, and
    then we have expanded the gradient descent law as
    \begin{multline}
        z_{1}\big(t_{\alphatime{p-1} + p}, \alphatime{p-1} + p\big) \\
        = \Pi_{\mathcal{U}}\bigg[z_0\big(t_{\alphatime{p-1} + p-1}, \alphatime{p-1} + p-1\big)
         \\ -\gamma\nabla_{\inpt} \Phi\Big(z_0\big(t_{\alphatime{p-1} + p-1}, \alphatime{p-1} + p-1\big), \\
         y_s\big(t_{\alphatime{p-1} + p-1}, \alphatime{p-1} + p-1\big)\Big)
        \Big)\bigg].
    \end{multline}
    Next, we have used the fact that the optimum is a fixed point
    of projected gradient descent, namely
    \begin{multline}
        \interm^*\big(t_{\alphatime{p-1} + p-1},\alphatime{p-1} + p-1\big) \\
        = \Pi_{\mathcal{U}}\Big[\interm^*\Big(t_{\alphatime{p-1} + p-1}, \alphatime{p-1} + p-1\big) \\ 
        - \gamma\nabla_u\Phi\Big(\interm^*\big(t_{\alphatime{p-1} + p-1}, \alphatime{p-1} + p-1\big), \\
        y_s\big(t_{\alphatime{p-1} + p-1}, \alphatime{p-1} + p-1\big)\Big)\Big], 
    \end{multline}
    and then applied the non-expansive property of~$\Pi_{\mathcal{U}}$
    to attain an upper bound by removing~$\Pi_{\mathcal{U}}$. 
    Then we have approximated the sampled output as
    described in Remark~\ref{rem:steadystate} via
    \begin{multline}
        y_s\big(t_{\alphatime{p-1} + p-1}, \alphatime{p-1} + p-1\big)\Big) \\
          = Hu(t_{\alphatime{p-1} + p-1}, \alphatime{p-1} + p-1) + d \\
          = Hz_0(t_{\alphatime{p-1} + p-1}, \alphatime{p-1} + p-1) + d.
    \end{multline}
    Next, combining like terms in~\eqref{eq:t1reallybigbound} and 
    applying the triangle inequality gives
     \begin{multline}
        \twonorm{\inpt{}\Hztime{\alphatime{p}+p}{\alphatime{p}+p} - \inpt{}^*\Hztime{\alphatime{p}+p}{\alphatime{p}+p}}\\
        \leq  q^{\frac{\alpha({p-1})-1}{2}} 
        \left\|
        z_0\Big(t_{\alphatime{p-1} + p-1}, \alphatime{p-1} + p-1\Big) \right. \\ 
        \left. - \interm^*\Big(t_{\alphatime{p-1} + p-1}, \alphatime{p-1} + p-1\Big) \right\| \\
        \cdot \|I_m - \gamma(Q_u + H^TQ_yH)\|.
    \end{multline}
    Using
    \begin{multline}
        \left\|
        z_0\Big(t_{\alphatime{p-1} + p-1}, \alphatime{p-1} + p-1\Big) \right. \\ 
        \left. - \interm^*\Big(t_{\alphatime{p-1} + p-1}, \alphatime{p-1} + p-1\Big) \right\| \leq d_{\mathcal{U}} 
    \end{multline}
    then gives the bound 
    \begin{multline} \label{eq:biguthing}
        \twonorm{\inpt{}\Hztime{\alphatime{p}+p}{\alphatime{p}+p} - \inpt{}^*\Hztime{\alphatime{p}+p}{\alphatime{p}+p}}\\
        \leq  q^{\frac{\alpha({p-1})-1}{2}} 
        d_{\mathcal{U}} \cdot \|I_m - \gamma(Q_u + H^TQ_yH)\|. 
    \end{multline}
    \blue{Onto Step~\ref{step:2},} using Lemma~\ref{lem:jpexconvg} \red{where $\tilde{x}$ is the optimal steady-state value}
    and adding and subtracting~$u^*(t_{\bar{\alpha}(P) + P}, \bar{\alpha}(P) + P)$ 
    we have 
    \begin{multline}
         \ste(t, j) - \tilde{x} 
        = e^{At}\ste\Hotime{0}{0} \\
        +\sum^{P-1}_{p=0}\int_{t_{\alphatime{p}+p}}^{t_{\alphatime{p+1}+p+1}}e^{A(t_{\alphatime{p+1}+p+1}-\intv)}d\intv\cdot\\
        B\big(\inpt\Hztime{\alphatime{p}+p}{\alphatime{p}+p} - \inpt{}^*\Hztime{\alphatime{p}+p}{\alphatime{p}+p}\big)\\
        + \int_{t_{\bar{\alpha}(P) + P}}^{t} e^{A(t - \tau)}d\tau \cdot \\
        B \big(u(t_{\bar{\alpha}(P) + P}, \bar{\alpha}(P) + P) - u^*(t_{\bar{\alpha}(P) + P}, \bar{\alpha}(P) + P)\big)\\
        +\sum^{P-1}_{p=0}\int_{t_{\alphatime{p}+p}}^{t_{\alphatime{p+1}+p+1}}e^{A(t_{\alphatime{p+1}+p+1}-\intv)}d\intv\cdot\\
        B\big(\inpt{}^*\Hztime{\alphatime{p}+p}{\alphatime{p}+p}\big) \\
        + \int_{t_{\bar{\alpha}(P) + P}}^{t} e^{A(t - \tau)}d\tau B u^*(t_{\bar{\alpha}(P) + P}, \bar{\alpha}(P) + P) -  \tilde{x}.
    \end{multline}
    Using $\tilde{x} = -A^{-1}B\tilde{u}$ and adding $-\tilde{u} + \tilde{u}$ gives
    \begin{multline} \label{eq:oneofsteps1}
         \ste(t, j) - \tilde{x} 
        = e^{At}\ste\Hotime{0}{0} \\
        +\sum^{P-1}_{p=0}\int_{t_{\alphatime{p}+p}}^{t_{\alphatime{p+1}+p+1}}e^{A(t_{\alphatime{p+1}+p+1}-\intv)}d\intv\cdot\\
        B\big(\inpt\Hztime{\alphatime{p}+p}{\alphatime{p}+p} - \inpt{}^*\Hztime{\alphatime{p}+p}{\alphatime{p}+p}\big)\\
        + \int_{t_{\bar{\alpha}(P) + P}}^{t} e^{A(t - \tau)}d\tau \cdot \\
        B \big(u(t_{\bar{\alpha}(P) + P}, \bar{\alpha}(P) + P) - u^*(t_{\bar{\alpha}(P) + P}, \bar{\alpha}(P) + P)\big)\\
        +\sum^{P-1}_{p=0}\int_{t_{\alphatime{p}+p}}^{t_{\alphatime{p+1}+p+1}}e^{A(t_{\alphatime{p+1}+p+1}-\intv)}d\intv\cdot\\
        B\big(\inpt{}^*\Hztime{\alphatime{p}+p}{\alphatime{p}+p} - \tilde{u}\big) \\
        + \int_{t_{\bar{\alpha}(P) + P}}^{t} e^{A(t - \tau)}d\tau B (u^*(t_{\bar{\alpha}(P) + P}, \bar{\alpha}(P) + P)  - \tilde{u}) \\
        +\sum^{P-1}_{p=0}\int_{t_{\alphatime{p}+p}}^{t_{\alphatime{p+1}+p+1}}e^{A(t_{\alphatime{p+1}+p+1}-\intv)}d\intv B \tilde{u} \\
        + \int_{t_{\bar{\alpha}(P) + P}}^{t} e^{A(t - \tau)}d\tau B \tilde{u}
        -  \tilde{x}.
    \end{multline}

    We know that
    \begin{multline}
        \sum^{P-1}_{p=0}\int_{t_{\alphatime{p}+p}}^{t_{\alphatime{p+1}+p+1}}e^{A(t_{\alphatime{p+1}+p+1}-\intv)}d\intv  \\
        + \int_{t_{\bar{\alpha}(P) + P}}^{t} e^{A(t - \tau)}d\tau = \int_{0}^{t} e^{A(t - \tau)} d\tau
    \end{multline}
    and that
    \begin{equation}
        \int_{0}^{t} e^{A(t - \tau)} d\tau = (e^{At} - I)A^{-1}. 
    \end{equation}
    These relations allow for simplifying \eqref{eq:oneofsteps1} to 
    \begin{multline}
      \ste(t, j) - \tilde{x}
        = e^{At}\ste\Hotime{0}{0} \\
        +\sum^{P-1}_{p=0}\int_{t_{\alphatime{p}+p}}^{t_{\alphatime{p+1}+p+1}}e^{A(t_{\alphatime{p+1}+p+1}-\intv)}d\intv\cdot\\
        B\big(\inpt\Hztime{\alphatime{p}+p}{\alphatime{p}+p} - \inpt{}^*\Hztime{\alphatime{p}+p}{\alphatime{p}+p}\big)\\
        + \int_{t_{\bar{\alpha}(P) + P}}^{t} e^{A(t - \tau)}d\tau \cdot \\
        B \big(u(t_{\bar{\alpha}(P) + P}, \bar{\alpha}(P) + P) - u^*(t_{\bar{\alpha}(P) + P}, \bar{\alpha}(P) + P)\big)\\
        +\sum^{P-1}_{p=0}\int_{t_{\alphatime{p}+p}}^{t_{\alphatime{p+1}+p+1}}e^{A(t_{\alphatime{p+1}+p+1}-\intv)}d\intv\cdot\\
        B\big(\inpt{}^*\Hztime{\alphatime{p}+p}{\alphatime{p}+p} - \tilde{u}\big) \\
        + \int_{t_{\bar{\alpha}(P) + P}}^{t} e^{A(t - \tau)}d\tau B (u^*(t_{\bar{\alpha}(P) + P}, \bar{\alpha}(P) + P)  - \tilde{u}) \\
        + e^{At}A^{-1}B\tilde{u}. 
    \end{multline}

    \blue{With Step~\ref{step:3},} we note that~$e^{At}x(0, 0) + e^{At}A^{-1}B\tilde{u} = e^{At}\big(x(0, 0) - \tilde{x}\big)$, 
    we take the norm of both sides and apply the triangle inequality to arrive at the bound
    \begin{multline}
      \twonorm{\ste(t, j) - \tilde{x}}
        = \twonorm{e^{At}} \twonorm{\ste\Hotime{0}{0} - \tilde{x}} \\
        +\sum^{P-1}_{p=0}\int_{t_{\alphatime{p}+p}}^{t_{\alphatime{p+1}+p+1}} \twonorm{e^{A(t_{\alphatime{p+1}+p+1}-\intv)}} d\intv\cdot\\
        \twonorm{B} \twonorm{\inpt\Hztime{\alphatime{p}+p}{\alphatime{p}+p} - \inpt{}^*\Hztime{\alphatime{p}+p}{\alphatime{p}+p}} \\
        + \int_{t_{\bar{\alpha}(P) + P}}^{t} \twonorm{e^{A(t - \tau)}} d\tau \cdot \\
        \twonorm{B} \twonorm{u(t_{\bar{\alpha}(P) + P}, \bar{\alpha}(P) + P) - u^*(t_{\bar{\alpha}(P) + P}, \bar{\alpha}(P) + P)} \\
        +\sum^{P-1}_{p=0}\int_{t_{\alphatime{p}+p}}^{t_{\alphatime{p+1}+p+1}} \twonorm{e^{A(t_{\alphatime{p+1}+p+1}-\intv)}} d\intv\cdot\\
        \twonorm{B} \twonorm{\inpt{}^*\Hztime{\alphatime{p}+p}{\alphatime{p}+p} - \tilde{u}} \\
        + \int_{t_{\bar{\alpha}(P) + P}}^{t} \twonorm{e^{A(t - \tau)}} d\tau \twonorm{B} \twonorm{u^*(t_{\bar{\alpha}(P) + P}, \bar{\alpha}(P) + P)  - \tilde{u}}. 
    \end{multline}
    Using \eqref{eq:biguthing} and the definition of~$d_{\mathcal{U}}$ gives
     \begin{multline} \label{eq:t1bigbound}
      \twonorm{\ste(t, j) - \tilde{x}}
        = \twonorm{e^{At}} \twonorm{\ste\Hotime{0}{0} - \tilde{x}} \\
        +\sum^{P-1}_{p=0}\int_{t_{\alphatime{p}+p}}^{t_{\alphatime{p+1}+p+1}} \twonorm{e^{A(t_{\alphatime{p+1}+p+1}-\intv)}} d\intv\cdot\\
        \twonorm{B}  q^{\frac{\alpha({p-1})-1}{2}} 
        d_{\mathcal{U}}\cdot\|I_m - \gamma(Q_u + H^TQ_yH)\|
        \\ + \int_{t_{\bar{\alpha}(P) + P}}^{t} \twonorm{e^{A(t - \tau)}} d\tau \cdot \\
        \twonorm{B}  q^{\frac{\alpha({p-1})-1}{2}} 
        d_{\mathcal{U}}\cdot\|I_m - \gamma(Q_u + H^TQ_yH)\| \\
        +\sum^{P-1}_{p=0}\int_{t_{\alphatime{p}+p}}^{t_{\alphatime{p+1}+p+1}} \twonorm{e^{A(t_{\alphatime{p+1}+p+1}-\intv)}} d\intv 
        \twonorm{B} d_{\mathcal{U}} \\
        + \int_{t_{\bar{\alpha}(P) + P}}^{t} \twonorm{e^{A(t - \tau)}} d\tau \twonorm{B} d_{\mathcal{U}}. 
    \end{multline}
    Assumption~\ref{as:timescale} ensures that~$\alpha(p) \geq \ell$
    for all~$p$.
    Using this fact and~$q \in (0,1)$, we have
    \begin{equation} \label{eq:t1qbound}
        q^{\frac{\alpha(p-1)-1}{2}} \leq q^{\frac{\ell-1}{2}}.
    \end{equation}
    Then using~\eqref{eq:t1qbound} in~\eqref{eq:t1bigbound} and combining sums of integrals into one integral as 
    was done below \eqref{eq:oneofsteps1} while separating out the $p=0$ case, we have
    \begin{multline} \label{eq:lastone}
      \twonorm{\ste(t, j) - \tilde{x}}
        = \twonorm{e^{At}} \twonorm{\ste\Hotime{0}{0} - \tilde{x}} \\                
        + \int_{0}^{t_{\alphatime{1} + 1}} \twonorm{e^{A(t_{\alphatime{1} + 1} - \tau)}} d\tau \twonorm{B}  
        d_{\mathcal{U}} \\
        + \int_{t_{\alphatime{1} + 1}}^{t} \twonorm{e^{A(t - \tau)}} d\tau \twonorm{B}  q^{\frac{\ell-1}{2}} 
        d_{\mathcal{U}} \\
        \cdot\|I_m - \gamma(Q_u + H^TQ_yH)\| \\
        + \int_{0}^{t} \twonorm{e^{A(t - \tau)}} d\tau \twonorm{B} d_{\mathcal{U}},
    \end{multline}
    %
    \blue{where the first integral does not have a~$q^{\frac{\ell}{2}}$ term to account for~$\alpha(-1)$.} Next, we observe that
    \begin{multline} \label{eq:t1bound1}
        \int_{0}^{t} \twonorm{e^{A(t-\intv)}} d\tau \leq M \int_{0}^{t} \exp\big({-\rho}(t - \tau) \big) d\tau \\
        = \frac{M}{\rho}\big[1 - \exp\big({-\rho} t\big)\big],
    \end{multline}
    where the inequality follows from Theorem~2 in~\cite[Chapter 1.9]{perko13}. 

   Substituting~\eqref{eq:t1bound1} into~\eqref{eq:lastone}, using
   \begin{equation} \label{eq:expNorm}
    \|e^{At}\| \leq M \exp({-\rho}t), 
   \end{equation}
   and rearranging terms gives 
    \begin{multline} 
        \twonorm{\ste(t, j) - \tilde{x}} 
        \leq M\exp({-\rho} t) \twonorm{\ste\Hotime{0}{0} - \tilde{x}} \\
        + \frac{M\twonorm{B} d_{\mathcal{U}}}{\rho} \big[1 - \exp({-\rho} t)\big]\\
        + \frac{M\twonorm{B} d_{\mathcal{U}} q^{\frac{\ell-1}{2}}}{\rho} \big[1 - \exp({-\rho} (t-t_{\alphatime{1} + 1}))\big]\cdot\\
        \|I_m - \gamma(Q_u + H^TQ_yH)\|\\
        + \frac{M\twonorm{B} d_{\mathcal{U}} q^{\frac{\ell-1}{2}}}{\rho} \big[1 - \exp({-\rho} t_{\alphatime{1} + 1})\big]\cdot\\
        \|I_m - \gamma(Q_u + H^TQ_yH)\|. 
    \end{multline}
    We find that 
    \begin{equation} \label{eq:sqrtqbound}
        \|I_m - \gamma(Q_u + H^TQ_yH)\| \leq q^{\frac{1}{2}}, 
    \end{equation}
    which follows from observing that
    \begin{multline} \label{eq:bignormbound} 
        \|I_m - \gamma(Q_u + H^TQ_yH)\|^2 \\ = \sup_{x : \|x\| = 1}
        \|[I_m - \gamma(Q_u + H^TQ_yH)]x\|^2 \\
        \leq 1 - \inf_{x : \|x\| = 1} 2\gamma x^T(Q_u + H^TQ_yH)x
        \\ + \gamma^2\|Q_u + H^TQ_yH\|^2. 
    \end{multline}
    Then, since~$H^TQ_yH \succeq 0$, we have
    \begin{multline} \label{eq:infbound}
        \inf_{x : \|x\| = 1} 2\gamma x^T(Q_u + H^TQ_yH)x \\ \geq \inf_{x : \|x\| = 1} 2\gamma x^TQ_ux
        = 2\gamma\lambda_{min}(Q_u).
    \end{multline}
    We use~\eqref{eq:infbound} and~\eqref{eq:L} 
    in~\eqref{eq:bignormbound},  
    and then taking the square root gives~\eqref{eq:sqrtqbound},
    from which we have the bound
    \begin{multline} \label{eq:C1lastone}
        \twonorm{\ste(t, j) - \tilde{x}} 
        \leq M\exp({-\rho} t) \twonorm{\ste\Hotime{0}{0} - \tilde{x}} \\
        + \frac{M\twonorm{B} d_{\mathcal{U}}}{\rho} \big[1 - \exp({-\rho} t)\big]\\
        + \frac{M\twonorm{B} d_{\mathcal{U}} q^{\frac{\ell}{2}}}{\rho} \big[1 - \exp({-\rho} (t-t_{\alphatime{1} + 1}))\big]\\
        + \frac{M\twonorm{B} d_{\mathcal{U}}}{\rho} \big[1 - \exp({-\rho} t_{\alphatime{1} + 1})\big].
    \end{multline}
    Next, using~$\atimer{,\min} \leq t_{\alphatime{1} + 1} \leq \atimer{,\max}$ and~$\exp({-\rho}(t - \atimer{,\min}) = \exp({\rho}\atimer{,\min})\exp({-\rho}t)$ gives the upper bound 
    \begin{multline}   \label{eq:C1lastoneone}
        \twonorm{\ste(t, j) - \tilde{x}} 
        \leq M\exp({-\rho} t) \twonorm{\ste\Hotime{0}{0} - \tilde{x}} \\
        + \frac{M\twonorm{B} d_{\mathcal{U}}}{\rho} \big[1 - \exp({-\rho} t)\big]\\
        + \frac{M\twonorm{B} d_{\mathcal{U}} q^{\frac{\ell}{2}}}{\rho} \big[1 -\exp({\rho}\atimer{,\min})\exp({-\rho}t)\big]\\
        + \frac{M\twonorm{B} d_{\mathcal{U}}}{\rho} \big[1 - \exp({-\rho} \atimer{,\max})\big].
    \end{multline}
    We observe that by definition of~$\|\cdot\|_{\mathcal{A}}$ we have
    \begin{equation} \label{eq:phi0relation}
        \|x(0, 0) - \tilde{x}\| \leq \|\phi(0, 0)\|_{\mathcal{A}}  + r. 
    \end{equation}
    Similarly, from~\eqref{eq:t1normsequal} we have 
    \begin{equation} \label{eq:phitjrelation}
        \|\phi(t, j)\|_{\mathcal{A}} = \|x(t, j)\|_{B_r(\tilde{x})} \leq \max\{0, \|x(t, j) - \tilde{x}\| - r\}. 
    \end{equation}
    Using~\eqref{eq:phi0relation} and~\eqref{eq:phitjrelation} in~\eqref{eq:C1lastoneone} gives
    \begin{multline} \label{eq:boundwithr}
        \twonorm{\phi(t, j)}_{\mathcal{A}} 
        \leq M\exp({-\rho} t) (\|\phi(0, 0)\|_{\mathcal{A}} + r)\\
        + \frac{M\twonorm{B} d_{\mathcal{U}}}{\rho} \big[1 - \exp({-\rho} t)\big]\\
        + \frac{M\twonorm{B} d_{\mathcal{U}} q^{\frac{\ell}{2}}}{\rho} \big[1 -\exp({\rho}\atimer{,\min})\exp({-\rho}t)\big]\\
        + \frac{M\twonorm{B} d_{\mathcal{U}}}{\rho} \big[1 - \exp({-\rho} \atimer{,\max})\big] - r,
    \end{multline} 
    where we can remove the~$\max$ operator that was in~\eqref{eq:phitjrelation} because the right-hand
    side of~\eqref{eq:boundwithr} is non-negative. 
    
    Then we combine like terms to find
    \begin{multline} \label{eq:t1end}
        \twonorm{\phi(t, j)}_{\mathcal{A}} \leq M\exp({-\rho} t) \|\phi(0, 0)\|_{\mathcal{A}} \\
        +\frac{M^2\twonorm{B}d_{\mathcal{U}}}{\rho}\Big(2 - \exp(-\rho\atimer{,\max}) + q^{\frac{\ell}{2}}\Big)\exp({-\rho} t)\\
        - \frac{M\twonorm{B}d_{\mathcal{U}}}{\rho}\Big(1 + q^{\frac{\ell}{2}}\exp({\rho}\atimer{,\min})\Big)\exp({-\rho} t)
    \end{multline} 

    We find the asymptotic behavior by taking the limit as $t+j$ goes to infinity, which implies that~$t$
    itself must go to infinity. 
    Then 
    \begin{equation} \label{eq:cor11}
        \lim_{t+j \to \infty} \exp(-\rho t) = 0. 
    \end{equation}    
    Using~\eqref{eq:cor11} in~\eqref{eq:t1end} gives~$\lim_{t + j \to \infty} \|\phi(t, j)\|_{\mathcal{A}} = 0$. 
 
\end{proof}}

\begin{remark} \label{rem:tunerho}
In the limit, the state~$x$ is asymptotically no farther
than a distance~$r$ from~$\tilde{x}$. The quantity~$r$ 
can be reduced 
with a pre-feedback controller that is applied before feedback
optimization is used.  
More precisely, if the pair~$(A, B)$ is controllable, then using pole placement we can 
use any~$\eta > 0$ to select $\rho$ such that 
\blue{$\rho \geq M\twonorm{B}d_{\mathcal{U}}\eta^{-1}(2 - \exp(-\rho\atimer{,\min})+ q^{\frac{\ell}{2}})$. }    
This~$\rho$ ensures~$r \leq \eta$ and hence that 
$\|x(t, j)\|_{B_{\eta}(\tilde{x})} \to 0$ for any desired~$\eta > 0$. 
\end{remark}


\subsection{Global Convergence and Robustness} \label{subsec:GConvandRobust}

Theorem \ref{th:completeHO} relies on~\eqref{eq:initconds}, which restricts 
it to only apply to initial conditions with~$\atimer{}(0,0) \in [\atimer{,min}, \atimer{,max}]$
and~$\ctimer{}(0,0) = \ctimer{,comp}$. 
We next derive a global convergence result that applies to solutions
that begin from arbitrary initial conditions, including
those that violate the conditions in~\eqref{eq:initconds}. 

\begin{theorem} [Global Complete Hybrid Convergence] \label{thm:GCHC}
    Consider the hybrid system~$\HFO$
    from~\eqref{eq:hybridFO} and suppose that  
    Assumptions \ref{LTI-AS} and~\ref{as:timescale} hold.
    Consider objectives of the form of~\eqref{eq:quadOBJ-AS}, and
    suppose that the gradient descent algorithm uses
    a stepsize~$\gamma \in \big(0,\frac{2}{\lambda_{min}(Q_u) + L}\big)$, 
    where~$Q_u$ is from~\eqref{eq:quadOBJ-AS} and~$L := \lambda_{max}(Q_u + H^{\top}Q_yH)$. 
    For each maximal solution~$\phi$ to~$\HFO$ with initial condition~$\phi(0, 0)$, for each~$(t, j) \in \textnormal{dom } \phi$,      
      \begin{multline} 
        \twonorm{\phi(t, j)}_{\mathcal{A}} \leq M\exp({-\rho} t) \|\phi(0, 0)\|_{\mathcal{A}} \\
        +\frac{M^2\twonorm{B}d_{\mathcal{U}}}{\rho}\Big(2 - \exp(-2\rho\atimer{,\max}) + q^{\frac{\ell}{2}}\Big)\exp({-\rho} t)\\
        - \frac{M\twonorm{B}d_{\mathcal{U}}}{\rho}\Big(1 + q^{\frac{\ell}{2}}\exp({\rho}\atimer{,\min})\Big)\exp({-\rho} t),
   \end{multline}
    where $d_{\mathcal{U}} = \max_{u_1, u_2 \in \mathcal{U}} \|u_1 - u_2\|$ is the diameter
    of the set~$\mathcal{U}$, $\rho$ is from~\eqref{eq:rhodef}, 
    $\ell \geq 1$ is from Assumption~\ref{as:timescale},
    $\mathcal{A}$ is from~\eqref{eq:closedsetA}, 
    \green{$q:= 1-2\gamma\lambda_{min}(Q_u) + \gamma^2L^2\in \paren{0,1}$}, 
    and~$M \geq 1$ is a constant. 
\end{theorem}

\ifbool{short}{
\begin{proof} 
\red{See authors' technical report~\cite{chuy2025hybridsystemsmodelfeedback}}
\end{proof}
}{
\begin{proof}
    To derive a bound without the conditions in~\eqref{eq:initconds}, we must consider 
    any~$\atimer{}(0, 0) \in [0, \atimer{,max}]$ and any~$\ctimer{}(0, 0) \in [0, \ctimer{,comp}]$. 
    The analysis in Theorem~\ref{th:completeHO} does not cover this case because it assumes
    that there are at least~$\ell \geq 1$ gradient descent iterations performed before
    the first change in the input, but that is no longer guaranteed
    if the conditions in~\eqref{eq:initconds} do not hold.     
    However, after a jump is triggered by~$\atimer{}$ reaching zero, 
    it is guaranteed that at least~$\ell$ gradient descent iterations will be performed
    before the second change in the input. 
    This second change in the input occurs at hybrid time
    $(t_{\bar{\alpha}(2) + 2}, \bar{\alpha}(2) + 2)$, and we focus
    on bounding the behavior of~$\HFO$ until this time. 
        
    The input over the interval~$[t_0, t_{\bar{\alpha}(1) + 1}]$
    is~$u(0, 0)$. To derive a bound that holds from all initial conditions,
    we consider~$\alpha(0) = 0$ gradient descent iterations
    being completed before time~$t_{\bar{\alpha}(1) + 1}$, which represents
    the least progress that the underlying optimization algorithm
    may make before the first change in the input. 
    Then the next input is simply equal to the previous input, and
    in particular
    the input over the interval~$[t_{\bar{\alpha}(1) + 1}, t_{\bar{\alpha}(2) + 2}]$
    is
    \begin{equation}
        u(t_{\bar{\alpha}(1) + 1}, \bar{\alpha}(1) + 1) = u(0, 0),
    \end{equation}
    which happens precisely because no computations have been performed to change the input. 
    This case is captured by
    using the same steps to reach~\eqref{eq:lastone} with~$\alpha(p) \geq \ell$ for all~$p \geq 2$, using~\eqref{eq:sqrtqbound}, and combining integrals we find
    \begin{multline} \label{eq:l4mainbound}
        \twonorm{\ste(t, j) - \tilde{x}}
        = \twonorm{e^{At}} \twonorm{\ste\Hotime{0}{0} - \tilde{x}} \\                
        + \int_{0}^{t_{\alphatime{2} + 2}} \twonorm{e^{A(t_{\alphatime{2} + 2} - \tau)}} d\tau \twonorm{B}  
        d_{\mathcal{U}} \\
        + \int_{t_{\alphatime{2} + 2}}^{t} \twonorm{e^{A(t - \tau)}} d\tau \twonorm{B}  q^{\frac{\ell}{2}} 
        d_{\mathcal{U}} \\
        + \int_{0}^{t} \twonorm{e^{A(t - \tau)}} d\tau \twonorm{B} d_{\mathcal{U}},
    \end{multline}
    where the first integral does not have a~$q^{\frac{\ell}{2}}$ term 
    to account the zero gradient descent iterations that are performed before the first jump in~$u$.

    Using~\eqref{eq:expNorm} in~\eqref{eq:l4mainbound} and integrating gives
    \begin{multline} 
        \twonorm{\ste(t, j) - \tilde{x}} 
        \leq M\exp({-\rho} t) \twonorm{\ste\Hotime{0}{0} - \tilde{x}} \\
        + \frac{M\twonorm{B} d_{\mathcal{U}}}{\rho} \big[1 - \exp({-\rho} t)\big]\\
        + \frac{M\twonorm{B} d_{\mathcal{U}} q^{\frac{\ell}{2}}}{\rho} \big[1 - \exp({-\rho} (t-t_{\alphatime{2} + 2}))\big]\\
        + \frac{M\twonorm{B} d_{\mathcal{U}}}{\rho} \big[1 - \exp({-\rho} t_{\alphatime{2} + 2})\big].
    \end{multline}
    Next, using~$\atimer{,\min} \leq t_{\alphatime{2} + 2} \leq 2\atimer{,\max}$ and~$\exp({-\rho}(t - \atimer{,\min}) = \exp({\rho}\atimer{,\min})\exp({-\rho}t)$ gives the upper bound

    \begin{multline} \label{eq:th2Lastone}
        \twonorm{\ste(t, j) - \tilde{x}} 
        \leq M\exp({-\rho} t) \twonorm{\ste\Hotime{0}{0} - \tilde{x}} \\
        + \frac{M\twonorm{B} d_{\mathcal{U}}}{\rho} \big[1 - \exp({-\rho} t)\big]\\
        + \frac{M\twonorm{B} d_{\mathcal{U}} q^{\frac{\ell}{2}}}{\rho} \big[1 -\exp({\rho}\atimer{,\min})\exp({-\rho}t)\big]\\
        + \frac{M\twonorm{B} d_{\mathcal{U}}}{\rho} \big[1 - \exp({-2\rho} \atimer{,\max})\big].
    \end{multline}
    The result follows from combining~\eqref{eq:phi0relation},~\eqref{eq:phitjrelation}, and~\eqref{eq:th2Lastone}. 
\end{proof}
}

In Theorem~\ref{thm:GCHC} one can 
use the value of~$\rho$ in Remark~\ref{rem:tunerho} to
force~$x$ to converge to a ball of any radius~$\eta > 0$
about~$\tilde{x}$.  

We can represent modeling errors as perturbations applied to the nominal system $\HFO$. 
We consider both errors in the LTI dynamics and 
errors in the timer dynamics. 
We first consider the perturbed domain
of the flow map, which is
\begin{multline} \label{eq:Crho}
C_\iota := \{\zeta \in \mathcal{X} : \atimer{}\in[0,\atimer{,\max} + \theta_{c,\max}], \\
        \ctimer{}\in[0,\ctimer{,comp} + \theta_{g,comp}]\},
\end{multline}
where~$\theta_{c,\max} \in (-\atimer{,\max}, \infty)$
and~$\theta_{g,comp} \in (-\ctimer{,comp}, \infty)$.
These perturbations allow~$\atimer{}$
to take values larger than~$\atimer{,\max}$ and similar
for~$\ctimer{}$. 
Although timers may be reset to inaccurate values, jumps are still triggered when at least one timer reaches zero and hence we use $D_\iota := D$ in the perturbed system model. 

In the flow map, there may be model errors in the~$A$ and~$B$
matrices, and the two timers may count down at a rate
that is not exactly~$1$. We define the perturbed
flow map as 
\begin{equation}
F_\iota(\zeta) :=
        \left(\begin{array}{cc}
            \big(A+\hat{A}\big)x + \big(B+\hat{B}\big)u \\   
                0 \\
                0 \\
                0 \\
            -1 + \kappa_{c} \\
            -1 + \kappa_{g}
        \end{array}\right), 
\end{equation}
where~$\hat{A} \in \mathbb{R}^{n \times n}$ models
errors in the~$A$ matrix, $\hat{B} \in \mathbb{R}^{n \times m}$
models errors in the~$B$ matrix, $\kappa_c \in (-\infty, 1)$
models errors in the rate at which~$\atimer{}$ counts down,
and~$\kappa_g \in (-\infty, 1)$ models errors in the rate
at which~$\ctimer{}$ counts down. 

Finally we define the perturbed jump map as 
    \begin{equation}
        G_\iota(\zeta) := 
        \begin{cases}
            G_{1,\iota}(\zeta) \quad \text{if }
            \atimer{} > 0 \textnormal{ and } \ctimer{} = 0 \quad \textnormal{ Case (i)} \\
            G_{2,\iota}(\zeta) \quad \text{if } \atimer{} = 0 \textnormal{ and } \ctimer{} > 0 \quad \textnormal{ Case (ii)} \\
            G_{3,\iota}(\zeta) \quad \text{if } \atimer{} = 0 \textnormal{ and } \ctimer{} = 0 \quad \textnormal{ Case (iii)}
        \end{cases}.
    \end{equation}
    For~$G_{1,\iota}$ we have
    \begin{equation}
        G_{1,\iota}(\zeta) := 
        \left(\begin{array}{cc}
            x\\
            u\\
            y_{s}\\
            \Pi_\mathcal{U}\left[z-\gamma\nabla_u\Phi(z, y_s)\right]\\
            \tau_{c} \\
            \tau_{g,comp} + \theta_{g,comp}
        \end{array}\right),
    \end{equation}
    where~$\theta_{g,comp}$ is from~\eqref{eq:Crho}. 
    This perturbed jump map allows for~$\ctimer{}$ to be reset
    to values other than~$\ctimer{,comp}$ after a gradient descent iteration
    is performed. 
    For~$G_{2,\iota}$ we have 
    \begin{equation}
        G_{2,\iota}(\zeta) := 
        \left(\begin{array}{cc}
            x\\
            z\\
            \big(H + \hat{H}\big)u + d\\
            z\\
            \left[\atimer{,\min} + \theta_{c,\min},\atimer{,\max} + \theta_{c,\max}\right]\\
            \tau_g
        \end{array}\right).
    \end{equation}
    Here,~$\hat{H} \in \mathbb{R}^{p \times m}$ models perturbations
    to the matrix~$H$, including those 
    that come from the errors~$\hat{A}$ and~$\hat{B}$ as described
    above, as well as errors in the output map~$\Psi$. 
    The interval to which~$\atimer{}$ is reset is perturbed
    with constants~$\theta_{c,\min} \in (-\atimer{,\min},\infty)$ and~$\theta_{c,\max} \in (-\atimer{,\max},\infty)$
    that satisfy
        $0 < \atimer{,\min} + \theta_{c,\min} \leq \atimer{,\max} + \theta_{c,\max}$,
    which ensures that~$\atimer{}$ is reset to a non-empty set, though both endpoints of the interval
    can be perturbed. 
    Finally we have
        $G_{3,\iota}(\zeta) := G_{1,\iota}(\zeta) \cup G_{2,\iota}(\zeta)$.
We define
\begin{multline} \label{eq:newrhodef}
\iota=\max\{
\theta_{g,comp},\|\hat{A}x\|,\|\hat{B}u\|,\|\hat{H}u\|, \\
\kappa_c,\kappa_g,\theta_{c,\min},\theta_{c,\max}\!
\}
\end{multline}

to be the maximum size of any perturbation at the state~$\zeta \in \mathcal{X}:= \R^{n + 2m + p + 2}$. 
The perturbed hybrid system model is 
\begin{equation} \label{eq:HFOR}
    \HFOR = 
   \begin{cases}
       \dot{\zeta} \in F_\iota(\zeta) & \zeta \in C_\iota\\
       \zeta^+\in G_\iota(\zeta) & \zeta \in D_\iota
   \end{cases}. 
\end{equation}

Our next theorem provides robustness guarantees for~$\HFO$. First, we require the following definition.

\ifbool{short}{
\begin{definition}[$(\tau, \epsilon)$-closeness {\cite[Definition 5.23]{Hybridbook}}] \label{def:teclose}
    Given $\tau,\epsilon > 0$, two hybrid arcs $\phi_1$ and $\phi_2$ are $(\tau,\epsilon)$-close if
    \blue{\begin{enumerate}
        \item for all $\!(t,\!j)\!\in\!\textnormal{dom }\phi_1$ with $t\!\!+\!\!j\!\leq\!\!\tau$ there exists $s$ such that $\!(s,\!j)\!\in\!\textnormal{dom }\phi_2$,$|t\!-\!s|\!<\!\epsilon$, and $\!|\phi_1(t,\!j)-\phi_2(s,j)|\!\!<\!\epsilon;$
        \item for all $\!(t,\!j)\!\in\!\textnormal{dom }\phi_2$ with $t\!\!+\!\!j\!\leq\!\!\tau$ there exists $s$ such that $\!(s,\!j)\!\in\!\textnormal{dom }\phi_1$,$|t\!-\!s|\!<\!\epsilon$, and $\!|\phi_2(t,\!j)-\phi_1(s,j)|\!\!<\!\epsilon$.
    \end{enumerate}}
\end{definition}
}{
\begin{definition}[$(\tau, \epsilon)$-closeness {\cite[Definition 5.23]{Hybridbook}}] \label{def:teclose}
    Given $\tau,\epsilon > 0$, two hybrid arcs $\phi_1$ and $\phi_2$ are $(\tau,\epsilon)$-close if
    \begin{enumerate}
        \item for all $(t,j)\in\textnormal{dom }\phi_1$ with $t+j\leq \tau$ there exists $s$ such that $(s,j)\in \textnormal{dom }\phi_2$, $|t-s|<\epsilon$, and 
        \begin{equation}
            |\phi_1(t,j)-\phi_2(s,j)|<\epsilon;
        \end{equation}
        \item for all $(t,j)\in\textnormal{dom }\phi_2$ with $t+j\leq \tau$ there exists $s$ such that $(s,j)\in \textnormal{dom }\phi_1$, $|t-s|<\epsilon$, and 
        \begin{equation}
            |\phi_2(t,j)-\phi_1(s,j)|<\epsilon.
        \end{equation}
    \end{enumerate}
\end{definition}
}
To the best of our knowledge, the next result is the first analytical characterization of the robustness of feedback optimization in a hybrid or sampled-data setting. 

\begin{theorem}[Robustness of $\HFO$]\label{th:robust} 
    Consider the hybrid system~$\HFOR$ with~$\iota$ as defined in~\eqref{eq:newrhodef}, and suppose that  
    Assumptions \ref{LTI-AS} and~\ref{as:timescale} hold. 
    Consider objectives of the form of~\eqref{eq:quadOBJ-AS}, and suppose that the gradient
    descent algorithm uses a stepsize~$\gamma \in \paren{0,\frac{2}{\lambda_{min}(Q_u) + L}}$. 
    Then, for every~$\epsilon > 0$ and~$\tau > 0$, there exists~$\delta > 0$
    with the following property: for every solution~$\phi_{\delta}$
    to~$\HFO^{\delta\iota}$, there exists a solution~$\phi$ to~$\HFO$ such
    that~$\phi_{\delta}$ and~$\phi$
    are~$(\tau,\epsilon)$-close. 
\end{theorem}

\ifbool{short}{
\begin{proof} 
\red{See authors' technical report~\cite{chuy2025hybridsystemsmodelfeedback}}
\end{proof}
}{
\begin{proof}
    \blue{Since $\HFO$ is well-posed and its maximal solutions are complete from all initial conditions, the result follows from~\cite[Proposition 6.34]{hybridfeedcntrl}.}
\end{proof}
}

\blue{
For perturbations in the model and timers, Theorem~\ref{th:robust} only holds over bounded hybrid time horizons, so the result is applied up until a chosen time~$\tau$. 
Then for some error~$\epsilon$, there is a nonzero perturbation~$\delta\iota$ of the perturbed system~$\HFO^{\delta\iota}$ such that its solution~$\phi_{\delta}$ is~$(\tau, \epsilon)$-close to the solutions of
the unperturbed system~$\HFO$. 
}


%
%
%
%

\section{Simulation Results} \label{sec:simulations}
\begin{figure*}[ht]
    \centering
    \includegraphics[width=.94\linewidth,keepaspectratio]{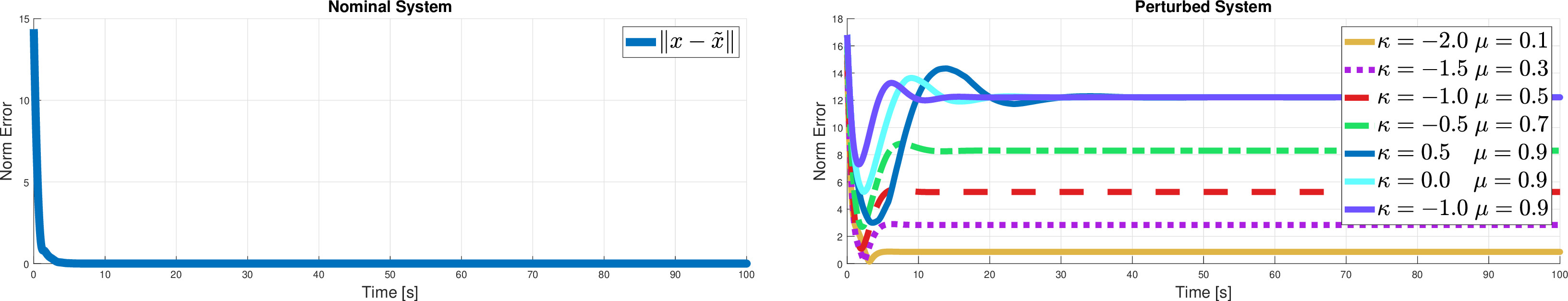}
    \caption{
    The left-hand plot shows the 
    convergence of the state~$x$ 
    under the dynamics
    $\mathcal{H}_{FO}$ from the initial condition in~\eqref{eq:initconds1}. 
    We see that~$x$
    attains an asymptotic error \blue{of~$3.26 \cdot 10^{-15}$, }
    which illustrates that it converges quite close to its desired value.
    The right-hand plot shows the convergence of the state~$x$
    under the perturbed dynamics~$\HFOR$ from the initial condition in~\eqref{eq:initconds1}.
    In this case, $x$ attains an asymptotic error \blue{of at most~$12.22$ relative to the errors from~$\HFO$
    under large model and timer perturbations.}
        \blue{
            under these perturbations, Assumption~\ref{LTI-AS} still guarantees exponential convergence.
        }
    }
    \label{fig:LSE-Combined}
\end{figure*}


In this section we present two sets of simulation results:
one with the nominal system~$\HFO$ and one
with the perturbed system~$\HFOR$. 
We consider the LTI dynamics 
\ifbool{short}{
\blue{\begin{equation}
        \dot{\ste} =
        \begin{bmatrix}
            -3 & 0 & 0 & 0\\
            3 & -3 & 0 & 0\\
            0 & 3 & -3 & 0\\
            0 & 0 & 3 & -3\\
        \end{bmatrix}
        \begin{bmatrix}
            x_1\\
            x_2\\
            x_3\\
            x_4
        \end{bmatrix}
        + \begin{bmatrix}
            3 \\
            0 \\
            0 \\
            0 \\
        \end{bmatrix}
        u \quad\quad
        \outpt = 
        x, 
\end{equation}}
}{
\begin{equation}
    \begin{aligned}
        \dot{\ste} &=
        \begin{bmatrix}
            -3 & 0 & 0 & 0\\
            3 & -3 & 0 & 0\\
            0 & 3 & -3 & 0\\
            0 & 0 & 3 & -3\\
        \end{bmatrix}
        \begin{bmatrix}
            x_1\\
            x_2\\
            x_3\\
            x_4
        \end{bmatrix}
        + \begin{bmatrix}
            3 \\
            0 \\
            0 \\
            0 \\
        \end{bmatrix}
        u \\
        \outpt &= 
        x, 
    \end{aligned}
\end{equation}
}
and
the feedback optimization problem we solve is
\begin{mini!}|l|
    {\inpt, \outpt{}}
    {\Phi(\inpt,y):=\frac{1}{2} Q_u u^2 + \frac{1}{2}(y-\hat{y})^\top Q_y (y - \hat{y})}
    {\label{probform2}}
    {}
    \addConstraint{\outpt = H\inpt + \dist, \,\, \inpt \in \mathcal{U}, \,\, \outpt{} \in \R^4}{}
\end{mini!}
where we use~$Q_u = 0.08$, $Q_y = 0.3I_4$, $\mathcal{U} = [-5,5]$, 
~$d = (0.2, 0.2, 0.2, 0.2)^\top$, and $\hat{y} = (4,4,4,4)^\top$. 

For simulations, the Hybrid Equations Toolbox (Version 3.0.0.76 \cite{Sanfelice_2013} ) was used, 
along with the initial conditions
\ifbool{short}{
\begin{multline} \label{eq:initconds1}
\blue{\ste(0,0)\!=\!(0,5,10,15)^\top, \;\inpt(0,0)\!=\! 0,~\atimer{}(0,0)\!=\!0.175,} \\
\soutpt{}(0,0)\!=\!\ste(0,0)\!+\!d,~\interm(0,0)\!=\!0, \;\ctimer{}(0,0)\!=\!0.05, 
\end{multline}
}{
\begin{multline} \label{eq:initconds1}
\ste(0,0)= (0,5,10,15)^\top, \inpt(0,0) = 0, \\ 
\soutpt{}(0,0) = (0.2,5.2,10.2,15.2)^\top, \interm(0,0)= 0, \\
\atimer{}(0,0) = 0.175,~\ctimer{}(0,0) = 0.05,
\end{multline}
}
where $\ctimer{,comp} = 0.05, \atimer{,\min} = 0.15 $, and $\atimer{,\max} = 0.20$ \blue{which ensures at least three iterations} with stepsize~$\gamma = 0.35 \leq \frac{2}{\lambda_{min}(Q_u)+L}$. 
We see in the left-hand plot
of Figure~\ref{fig:LSE-Combined} that $\|x-\tilde{x}\|$ 
converges \blue{to~$3.26 \cdot 10^{-15}$,} and thus~$x$ is 
asymptotically close to its desired value. 

We next consider the perturbed case.
Let~$J_{a, b}$ denote the matrix
of all ones in~$\mathbb{R}^{a \times b}$. 
For the timers, we use perturbations of the form
\blue{$\theta_{c,\min} = \theta_{c,\max} = \theta_{g, comp} = 0.5$ and~$\kappa_c = \kappa_g = \kappa$
for several values of~$\kappa$,} and for the LTI dynamics 
we use perturbations of the form~$\hat{A} = \mu J_{n\times n}$,
$\hat{B} = \mu J_{n \times m}$,
and~$\hat{H} = \mu J_{p \times m}$ for several values of~$\mu$.
The values of~$\kappa$ and~$\mu$ used in simulations are shown in Table~\ref{tab:perturbations},
along with the values they induce in the steady-state error~$\|x - \tilde{x}\|$  \blue{(from~$0.86$ to~$12.22$)}. 
Using these values, 
the perturbation~$\rho$ from~\eqref{eq:newrhodef}
can be quite large because
it is the maximum of several terms,
including~$\|\hat{A}x\|$, $\|\hat{B}u\|$,
and~$\|\hat{H}u\|$, which
can be large because they depend on~$x$ or~$u$. 
\blue{A second-order regression takes the form
~$\|\mathbf{x} - \tilde{\mathbf{x}}\|\approx 0.217 + 6.054\mu + 8.036\mu^2$,
showing that the steady-state error is dependent on the model perturbations rather than the time errors. This agrees with the right-hand plot of Figure~\ref{fig:LSE-Combined} where $\kappa$ affected the convergence rate. 
}


\begin{table}
\centering
\caption{Values of perturbations and
the asymptotic error for the perturbed system~$\HFOR$ relative to $\HFO$.}
\begin{tabular}{c|ccccc|}
\cline{2-6}  
& \multicolumn{5}{c|}{Value of $\|x - \tilde{x}\|$} \\ 
\cline{1-6}
\multicolumn{1}{|c|}{$\kappa $}    & \multicolumn{1}{c|}{-2.0} & \multicolumn{1}{c|}{-1.5} & \multicolumn{1}{c|}{-1.0} & \multicolumn{1}{c|}{-0.5} & \multicolumn{1}{c|}{0.5} \\ 
\hline\hline
\multicolumn{1}{|c|}{$\mu = 0.1$} & \multicolumn{1}{c|}{0.86} & \multicolumn{1}{c|}{0.86} & \multicolumn{1}{c|}{0.86} & \multicolumn{1}{c|}{0.86} & \multicolumn{1}{c|}{0.86}            
\\ \hline
\multicolumn{1}{|c|}{$\mu = 0.3$} & \multicolumn{1}{c|}{2.84}  & \multicolumn{1}{c|}{2.84} & \multicolumn{1}{c|}{2.84} & \multicolumn{1}{c|}{2.84} &  2.84           
\\ \hline
\multicolumn{1}{|c|}{$\mu = 0.5$} & \multicolumn{1}{c|}{5.26} & \multicolumn{1}{c|}{5.26} & \multicolumn{1}{c|}{5.26} & \multicolumn{1}{c|}{5.26} & 5.26             
\\ \hline
\multicolumn{1}{|c|}{$\mu = 0.7$} & \multicolumn{1}{c|}{8.30} & \multicolumn{1}{c|}{8.30} & \multicolumn{1}{c|}{8.30} & \multicolumn{1}{c|}{8.30} &  8.30            
\\ \hline
\multicolumn{1}{|c|}{$\mu = 0.9$} & \multicolumn{1}{c|}{12.22} & \multicolumn{1}{c|}{12.22} & \multicolumn{1}{c|}{12.22} & \multicolumn{1}{c|}{12.22} &  12.22       
\\ \hline
\end{tabular}\\
\label{tab:perturbations}
\end{table}

\section{Conclusion} \label{sec:conclusion}
This paper presented a hybrid system model for feedback optimization that considers continuous-time dynamics with discrete-time optimization. 
We showed that its maximal solutions are complete and non-Zeno, and then
we bounded their distance to a desired goal state. 
We also presented what are, to the best of our knowledge, the first analytical robustness
results for feedback optimization in a hybrid/sampled-data setting. 
Future work includes using nonconvex objective functions and using hybrid feedback optimization for systems with nonlinear dynamics.

\bibliographystyle{IEEEtran}{}
\bibliography{Bib}

\ifbool{short}{}{\appendix
\subsection{Outer Semicontinuity of Jump Maps} \label{app:gosc}
\begin{lemma}[{\cite[Lemma A.33]{hybridfeedcntrl}}] \label{OSLBcond}
     Given closed sets $D_1\subset \R^m$ and $D_2\subset \R^m$ and the set-valued maps $G_1:D_1 \rightrightarrows \R^n$ and $G_2:D_2\rightrightarrows\R^n$ that are outer semicontinuous and locally bounded relative to $D_1$ and $D_2$, respectively, the set-valued map $G:D\rightrightarrows \R^n$ given by
    \begin{align}
    G(\Stes) &:= G_1(\Stes) \cup G_2(\Stes) \\
             &= 
        \begin{cases}
            G_1(\Stes) &\text{if } \Stes \in D_1 \backslash D_2 \\
            G_2(\Stes) &\text{if } \Stes \in D_2 \backslash D_1\\
            G_1(\Stes)\cup G_2(\Stes) &\text{if } \Stes \in D_1 \cap D_2
        \end{cases}\label{OSLBjump}
    \end{align}
    for each $\Stes \in D$ is outer-semicontinuous and locally bounded relative to the closed set $D$. 
\end{lemma}

\subsection{Completeness of Maximal Solutions} \label{app:comp}
\begin{lemma}[Basic existence of solutions; Proposition~2.34 in \cite{hybridfeedcntrl}] \label{lem:complete}
     Let $\mathcal{H}=(C,F,D,G)$ satisfy Definition~\ref{def:hybridcond}. Take an arbitrary $\nu \in C \cup D.$ If $\nu \in D$ or \\
    (VC) there exists a neighborhood $U$ of $\nu$ such that for every $\zeta \in U \cap C$,
    \begin{equation*}
        F(\zeta) \cap T_C(\zeta) \neq \emptyset, 
    \end{equation*}
    then there exists a nontrivial solution $\phi$ to $\mathcal{H}$ with $\phi(0,0)=\nu$. 
    If (VC) holds for every $\nu \in C\backslash D$, then there exists a nontrivial solution to $\mathcal{H}$ from every initial point in $C \cup D$, and every maximal solution~$\phi$ 
    to~$\mathcal{H}$ 
    satisfies exactly one of the following conditions:
    \begin{enumerate}
        \item $\phi$ is complete;
        \item dom $\phi$ is bounded and the interval $I^J$, where $J = \sup_j \textnormal{dom } \phi$, has nonempty interior and $t\mapsto \phi(t,J)$ is a maximal solution to $\dot{z}\in F(z)$, in fact lim$_{t\rightarrow T}|\phi(t,J)|=\infty$, where $T = \sup_t \textnormal{dom } \phi$;
        \item $\phi(T,J)\in C\cup D $, where $(T,J)= \sup \textnormal{dom } \phi$. 
    \end{enumerate}
    Furthermore, if $G(D)\subset C \cup D$, then $3)$ above does not occur. 
\end{lemma} }

\end{document}